\documentclass[12pt]{article}
\usepackage{epsfig}
\usepackage{psfrag}
\usepackage{latexsym}
\usepackage[DIV13]{typearea}
\usepackage{amsmath}
\usepackage{amssymb}
\usepackage{amsfonts}
\usepackage{cite}
\usepackage{bbold}
\usepackage[footnotesize]{caption2}
\usepackage{graphicx}
\usepackage[center,footnotesize,hang]{subfigure}
\usepackage{url}
 \def\cO{{\cal O}}

\newcommand{\phit}{\varphi_T}

\newcommand{\phis}{\varphi_S}

\newcommand{\xit}{\tilde{\xi}}

\newcommand{\xipp}{\xi^{\prime\prime}}

\newcommand{\vt}{v_\text{T}}

\newcommand{\Uu}{\text{U(1)}}

\newcommand{\SUd}{\text{SU(2)}}

\newcommand{\Zt}{\text{Z}_3}

\def\gs{\mathrel{
   \rlap{\raise 0.511ex \hbox{$>$}}{\lower 0.511ex \hbox{$\sim$}}}}
\def\ls{\mathrel{
   \rlap{\raise 0.511ex \hbox{$<$}}{\lower 0.511ex \hbox{$\sim$}}}}

\newcommand{\ba}{\begin{array}{c}}
\newcommand{\baz}{\begin{array}{cc}}
\newcommand{\bad}{\begin{array}{ccc}}
\newcommand{\ea}{\end{array}}

\newcommand{\be}{\beta}

\newcommand{\La}{\Lambda}
\newcommand{\om}{\omega}

\newcommand{\mean}[1]{\langle#1\rangle}

\newcommand{\onep}{1^\prime}
\newcommand{\onepp}{1^{\prime\prime}}
\newcommand{\twop}{2^\prime}
\newcommand{\twopp}{2^{\prime\prime}}

\def\beq{\begin{equation}}
\def\eeq{\end{equation}}
\def\bea{\begin{eqnarray}}
\def\eea{\end{eqnarray}}
\def\bet{\begin{tabular}}
\def\eet{\end{tabular}}
\def\bes{\begin{subequations}\bea}
\def\ees{\eea\end{subequations}}

\def\be{\begin{equation}}
\def\ee{\end{equation}}
\def\bc{\begin{center}}
\def\ec{\end{center}}
\def\bea{\begin{eqnarray}}
\def\eea{\end{eqnarray}}
\def\dd{\displaystyle}
\def\nn{\nonumber}

\catcode`@=11
\def\marginnote#1{}
\newcount\hour
\newcount\minute
\newtoks\amorpm
\hour=\time\divide\hour by60
\minute=\time{\multiply\hour by60 \global\advance\minute by-\hour}
\edef\standardtime{{\ifnum\hour<12 \global\amorpm={am}%
        \else\global\amorpm={pm}\advance\hour by-12 \fi
        \ifnum\hour=0 \hour=12 \fi
        \number\hour:\ifnum\minute<10 0\fi\number\minute\the\amorpm}}
\edef\militarytime{\number\hour:\ifnum\minute<10 0\fi\number\minute}
\def\draftlabel#1{{\@bsphack\if@filesw {\let\thepage\relax
   \xdef\@gtempa{\write\@auxout{\string
      \newlabel{#1}{{\@currentlabel}{\thepage}}}}}\@gtempa
   \if@nobreak \ifvmode\nobreak\fi\fi\fi\@esphack}
        \gdef\@eqnlabel{#1}}
\def\@eqnlabel{}
\def\@vacuum{}
\def\draftmarginnote#1{\marginpar{\raggedright\scriptsize\tt#1}}
\def\draft{\oddsidemargin 0.0truein
        \def\@oddfoot{\sl preliminary draft \hfil
        \rm\thepage\hfil\sl\today\quad\militarytime}
        \let\@evenfoot\@oddfoot \overfullrule 3pt
        \let\label=\draftlabel
        \let\marginnote=\draftmarginnote
   \def\@eqnnum{(\theequation)\rlap{\kern\marginparsep\tt\@eqnlabel}%
\global\let\@eqnlabel\@vacuum}  }
\catcode`@=12
\begin{document}
\begin{titlepage}
\vspace*{0cm}
\phantom{hep-ph/***} 
\hfill{DFPD-07/TH/02}
\vskip 2.5cm
\begin{center}
{\Large\bf Tri-bimaximal Neutrino Mixing and Quark Masses\\ 
\vskip .3cm
from a Discrete Flavour Symmetry}
\end{center}
\vskip 0.5  cm
\begin{center}
{\large Ferruccio Feruglio}~$^{a)}$\footnote{e-mail address: feruglio@pd.infn.it},
{\large Claudia Hagedorn}~$^{b)}$\footnote{e-mail address: hagedorn@mpi-hd.mpg.de},
\\
\vskip .2cm
{\large Yin Lin}~$^{a)}$\footnote{e-mail address: yin.lin@pd.infn.it} and
{\large Luca Merlo}~$^{a)}$\footnote{e-mail address: merlo@pd.infn.it}
\\
\vskip .2cm
$^{a)}$~Dipartimento di Fisica `G.~Galilei', Universit\`a di Padova 
\\ 
INFN, Sezione di Padova, Via Marzolo~8, I-35131 Padua, Italy
\\
\vskip .1cm
$^{b)}$~
Max-Planck-Institut f\"ur Kernphysik
\\ 
Postfach 10 39 80, 69029 Heidelberg, Germany
\end{center}
\vskip 0.7cm
\begin{abstract}
\noindent
We build a supersymmetric model of quark and lepton masses based on the discrete flavour symmetry group $T'$, the double covering of $A_4$.
In the lepton sector our model is practically indistinguishable from recent models based on $A_4$ and, in particular, it predicts
a nearly tri-bimaximal mixing, in good agreement with present data. In the quark sector a realistic pattern of masses and mixing angles
is obtained by exploiting the doublet representations of $T'$, not available in $A_4$. To this purpose, the flavour symmetry $T'$ should be 
broken spontaneously along appropriate directions in flavour space. In this paper we fully discuss the related vacuum alignment problem,
both at the leading order and by accounting for small effects coming from higher-order corrections. As a result we get the relations:
$\sqrt{m_d/m_s}\approx \vert V_{us}\vert$ and $\sqrt{m_d/m_s}\approx \vert V_{td}/V_{ts}\vert$.
\end{abstract}
\end{titlepage}
\setcounter{footnote}{0}
\vskip2truecm
\section{Introduction}
It is a remarkable fact that, despite an intense and continuous theoretical effort over many years, our understanding of fermion masses
and mixing angles remains at a very primitive level. Our theoretical constructions clearly suffer from the lack of a guiding principle
and we can only replicate variants of very few basic ideas. Perhaps one of the most fruitful ideas in the field is the one by Froggatt and Nielsen 
(FN) \cite{fn},
who advocated a spontaneously broken flavour symmetry to account for the small mass ratios and the small mixing angles characterizing 
the quark sector. This idea was originally formulated with a U(1) flavour symmetry group, broken by the vacuum expectation value (VEV)
of a single scalar field, but it has been subsequently extended and adapted to a large variety of symmetry groups, with more elaborated
symmetry breaking patterns. While the original U(1) flavour symmetry is still a viable candidate to reproduce, at least at the order-of-magnitude
level, both quark and lepton masses and mixing angles \cite{reviewaf}, models based on the discrete symmetry group $A_4$ \cite{ma1,ma2,af1,af2,afl}
have been recently singled
out as good candidates to explain the approximate tri-bimaximal (TB) mixing \cite{hps} observed in the lepton sector. As a matter of fact
the range for the lepton mixing angles, as determined from a global fit to neutrino oscillation data, is given by (2$\sigma$ errors) \cite{data}:
\begin{equation}
\sin^2\theta_{12}=0.314(1^{+0.18}_{-0.15})~~~,~~~~~
\sin^2\theta_{23}=0.45(1^{+0.35}_{-0.20})~~~,~~~~~
\sin^2\theta_{13}=0.8^{+2.3}_{-0.8}\times 10^{-2}~~~,
\label{nma}
\end{equation}
and it is perfectly compatible with a TB mixing:
\begin{equation}
\sin^2\theta_{12}^{TB}=\dd\frac{1}{3}~~~,~~~~~
\sin^2\theta_{23}^{TB}=\dd\frac{1}{2}~~~,~~~~~
\sin^2\theta_{13}^{TB}=0
\label{tbm}
\end{equation}
Given the present experimental uncertainties, the agreement is not so impressive as far as $\theta_{23}$ and $\theta_{13}$ are concerned,
but it is certainly striking for the angle $\theta_{12}$:
\be
\theta_{12}^{TB}=35.3^0~~~~~~~~~~~~~~\theta_{12}=(34.1^{+1.7}_{-1.6})^0~~~(1\sigma)
\ee
given the small 1$\sigma$ error, less than $0.04~{\rm rad}\approx\lambda^2$, $\lambda\approx 0.22$ being the Cabibbo angle.
Measurements in a near future will bring down the precision/sensitivity on $\theta_{23}$ and $\theta_{13}$ to a similar level \cite{schwetz} , thus providing  
a stringent test of the TB mixing scheme.

It is well-known by now that, by working in the framework of a FN model, the only way to achieve TB lepton mixing is to set up an appropriate
vacuum alignment mechanism. The starting point is a flavour symmetry group $F$, spontaneously broken by the VEVs of a set of scalar fields.
Neutrino masses and charged lepton masses should be mainly sensitive to two separate sets of scalar fields, whose VEVs 
break the symmetry $F$ down to two different subgroups.
This VEV misalignment can produce the TB mixing, at the leading order. A complete separation between
VEVs affecting neutrino masses and charged lepton masses is however impossible, since a mixing is generally induced at some level by 
higher dimensional operators allowed by the symmetries of the model. Such a mixing can be kept under control if the typical expansion
parameters, the dimensionless ratios between the VEVs and the cut-off of the theory, are sufficiently small. Along these lines it is possible
to construct a model reproducing the TB mixing, perturbed by small corrections, below the $\lambda^2$ level.

Perhaps the simplest examples of this kind of models are those based on the flavour symmetry group $A_4$, 
the group of even permutations of four objects, also equal to the group of proper rotations leaving a regular tetrahedron invariant. This group is small,
it has only 12 elements and four inequivalent irreducible representations: a triplet one and three independent singlets $1$, $1'$ and $1''$. Lepton SU(2)
doublets $l_i$ $(i=e,\mu,\tau)$ are assigned to the triplet $A_4$ representation, while the lepton singlets $e^c$, $\mu^c$ and $\tau^c$ are assigned to $1$, $1''$ and $1'$, respectively. The symmetry breaking sector consists of scalar fields neutral under the SM gauge group: $(\varphi_T,\varphi_S,\xi)$,
transforming as $(3,3,1)$ of $A_4$.
Additional ingredients are needed in order to reproduce the desired alignment. The simplest version of the model is supersymmetric (though
SUSY is not really necessary to achieve the alignment) and possesses an additional $Z_3$ discrete symmetry that eliminates unwanted 
operators. The key feature of the model is that the minimization of the scalar potential at the leading order, that is by neglecting higher dimensional
operators, leads to the following VEVs:
\begin{equation}
\langle \varphi_T\rangle\propto (1,0,0)~~~,~~~~~\langle \varphi_S\rangle\propto (1,1,1)~~~,~~~~~\langle \xi\rangle\ne 0~~.
\label{va1}
\end{equation}
In the basis chosen for the generators of $A_4$, these VEVs imply a diagonal mass matrix in the charged lepton sector. 
In this sector the relative hierarchy between $m_e$, $m_\mu$ and $m_\tau$
can be controlled by a FN U(1)$_{FN}$ flavour symmetry. At the same time, the neutrino mass matrix gives rise to the TB mixing, independently 
from the values of the free parameters, which, in a finite portion of the parameter space, only affect the neutrino mass eigenvalues.
Of course, while the specific form of neutrino and charged lepton mass matrices does depend on the basis chosen, the physical properties
of the system, such as the mass eigenvalues and the mixing angles, are basis independent features.

It would be desirable to extend this construction to the quark sector, thus realizing a coherent description of all fermion masses,
but the simplest extrapolations explored so far turn out to be unrealistic. In the simplest possible extension,
quark SU(2) doublets $q_i$ are assigned to the triplet $A_4$ representation, while the quark SU(2) singlets $(u^c,d^c)$, $(c^c,s^c)$  
and $(t^c, b^c)$ are assigned to $1$, $1''$ and $1'$, respectively. Then, given the VEVs in eq. (\ref{va1}), 
and taking into account the additional $Z_3$ symmetry, at the leading order the quark mass matrices
in the up and down sectors are both diagonal and the quark mixing matrix $V_{CKM}$ is the unity matrix. 
Subleading corrections, coming from higher-dimensional operators contributing to quark masses were analyzed in ref. \cite{af2} and
are too small. A possible way out might be to consider new sources of symmetry breaking. This can consist in an explicit breaking
\cite{maq}, which however does not allow a complete control of the model and introduces a high degree of arbitrariness. 
Otherwise it can be realized by extending the symmetry breaking sector,
by allowing for some new scalar fields, whose VEVs could substantially contribute to the quark sector, giving rise to a realistic Cabibbo angle
$\lambda$. The difficulty with such an option is that these new scalar fields tend to affect also the lepton sector, giving rise to too
large, unacceptable, corrections to the TB mixing pattern. 
Finally, a disturbing feature of this construction is that the mass of the top quark comes from a non-renormalizable operator,
as for all the other fermions. This is against expectation, since in the SM the Yukawa coupling of the top quark
is of order one. To reproduce such a Yukawa coupling from a non-renormalizable operator we should introduce
large dimensionless couplings, which is unnatural. A discussion about the symmetry breaking pattern suitable to produce
realistic mixing angles in the quark sector can be found in ref. \cite{volkas}.

In the present paper we explore a different possibility, by considering as a starting point of our FN construction, the $T'$ group, the double
covering of $A_4$. The relation between $T'$ and $A_4$ is quite similar to the familiar relation between SU(2) and SO(3).
In particular,  SU(2) and SO(3) possess the same Lie algebra, but SO(3) has only integer representations,
while SU(2) possesses both integer and half-integer representations. Similarly, the representations of $T'$ are those of $A_4$
plus three independent doublets $2$, $2'$ and $2''$. By working only with the triplet and singlet representations, $T'$ is indistinguishable
from $A_4$. This allows to replicate with $T'$ the successful construction realized within $A_4$ in the lepton sector.
At the same time, the presence of the doublet representations can be exploited to describe the quark sector.
It is natural to assign quarks to a reducible singlet plus doublet representation. This has several advantages. By assigning the third
generation to a singlet representation the top quark can acquire mass already at the renormalizable level.
Moreover, by using doublets to describe quarks of first two generations, the VEVs in eq. (\ref{va1}) provide masses for the charm and the strange
quarks, which are conveniently suppressed with respect to the top and bottom masses. The mixing between second and third generations
can be induced by the VEV of a $T'$ scalar doublet, whose effects do not modify the TB mixing in the lepton sector.
First generation masses and mixing angles can arise through subleading effects. Actually, the whole picture in the quark sector
is very similar to that detailed in a series of papers \cite{aranda} exploiting the $T'$ group and it is also very close to that emerging from the study of another
popular FN group, U(2) \cite{barbieri}. In our paper we will combine the good known features of the old U(2) and $T'$ constructions for the quark sector,  
with the ability of $T'$ in reproducing the TB mixing in the lepton sector \cite{carr}. Such a combination is far from trivial and is not exhausted
by a list of particle representations. On the contrary, the key point is represented by the study of the vacuum alignment problem.
With an enlarged symmetry breaking sector we will analyze, both at the leading and at the subleading level the symmetry breaking
pattern of $T'$ and we will explicitly check the existence of a finite portion of the parameter space that gives rise to TB lepton mixing
and to a realistic pattern of quark masses and mixing angles. 
Eventually we get a very appealing group theoretical interpretation of the difference between quark and lepton mixing angles.
Large lepton mixing angles correspond to a breaking of the flavour symmetry group down to two different subgroups in the neutrino and in the charged lepton sectors.
Small quark mixing angles arise from the breaking of the flavour group along the same subgroup both in the up and in the down sectors.
In our model the gauge group is that of the Standard Model. For recent attempts
to incorporate quarks in a grand unified picture with flavour symmetry $A_4$, see ref. \cite{a4unified}.
%
%
\section{The group $T'$}
Our model is based on the flavour group $F=T'\times ...$ where $T'$ is the binary tetrahedral group \cite{tprime} that we will describe in this section and 
dots denote some additional group factor that we will specify later on. The key role in our construction is played by the $T'$ group that is literally
the double covering of the tetrahedral group $A_4$. The relation between $T'$ and $A_4$ can be understood by thinking of $A_4$, the group
of proper rotation in the three-dimensional space leaving a regular tetrahedron invariant, as a subgroup of SO(3). Thus the 12 elements of
$A_4$ are in a one-to-one correspondence with 12 sets of Euler angles. Now consider SU(2), the double covering of
SO(3), possessing ``twice'' as many elements as SO(3). There is a correspondence from SU(2) to SO(3) that maps two distinct elements of SU(2)
into the same set of Euler angles of SO(3). The group $T'$ can be defined as the inverse image under this map of the group $A_4$.

 The group $T'$ has 24 elements and has two kinds of representations. It contains the representations of $A_4$: one triplet 3 and three singlets $1$, $1'$ and $1''$.
 When working with these representations there is no way to distinguish the group $T'$ from the group $A_4$. In particular, in these representations, the elements of
 $T'$ coincide two by two and can be described by the same matrices that represent the elements in $A_4$.
 The other representations are three doublets $2$, $2'$ and $2''$. 
 The representations $1^{\prime}$, $1^{\prime \, \prime}$ and $2 ^{\prime}$, $2 ^{\prime \, \prime}$ are complex conjugated
to each other. 
Note that $A_{4}$ is not a subgroup of $T^{\prime}$, since the two-dimensional representations cannot be decomposed into representations of $A_{4}$.
The character table is shown in Table \ref{chartabTprime}.
\begin{table}
\begin{center}
\begin{tabular}{l|ccccccc|}
&\multicolumn{7}{|c|}{classes}                                                  \\ \cline{2-8}
&$C_{1}$&$C_{2}$&$C_{3}$&$C_{4}$&$C_{5}$&$C_{6}$&$C_{7}$\\
\cline{1-8}
\rule[0.15in]{0cm}{0cm} $\rm T$  &$\rm E$&$\rm \mathbb{R}$&$\rm C _{2}$,$\rm C _{2} \mathbb{R}$&$\rm C _{3}$&$\rm C^{2} _{3}$&$\rm C _{3} \mathbb{R}$&$\rm C^{2} _{3} \mathbb{R}$\\
\cline{1-8}
\rule[0.15in]{0cm}{0cm} $\rm G$  &$\rm \mathbb{1}$&$\rm \mathbb{R}$&$S$&$S T \mathbb{R}$&$T^{2}$&$T$&$(S T)^{2} \mathbb{R}$\\
\cline{1-8}
$^{\circ} C_{i}$                   &1      &1      &6      &4      &4      &4      &4\\
\cline{1-8}
$^{\circ} h_{C_{i}}$               &1      &2      &4      &6     &3      &3       &6\\
\hline
$1$                                &1      &1      &1      &1     &1      &1       &1 \\
$1 ^{\prime}$                      &1      &1      &1    &$\omega$    &$\omega^{2}$      &$\omega$    &$\omega^{2}$\\
$1 ^{\prime \, \prime}$            &1      &1      &1    &$\omega ^{2}$    &$\omega$     &$\omega ^{2}$    &$\omega$\\
$2$                                &2      &-2     &0    &1       &-1      &-1         &1\\
$2 ^{\prime}$                      &2      &-2     &0    &$\omega$         &$-\omega^{2}$   &$-\omega$   &$\omega ^{2}$\\
$2 ^{\prime \, \prime}$            &2      &-2     &0    &$\omega ^{2}$   &$-\omega$    &$-\omega^{2}$       &$\omega$\\
$3$                                &3      &3      &-1   &0      &0        &0         &0\\
\end{tabular}
\end{center}
\begin{center}
\begin{minipage}[t]{12cm}
\caption[]{Character table of the group
  $T^{\prime}$ taken from \cite{aranda}. $\omega$  is the third root of unity, i.e. $\omega= e^{\frac{2 \pi i}{3}} =  -\frac{1}{2} + i \frac{\sqrt{3}}{2}$.  $C_{i}$ are the classes of the
group, $^{\circ} C_{i}$ is the order of the $i ^{\mathrm{th}}$ class, i.e. the number of distinct elements contained in this class, $^{\circ} h_{C_{i}}$
is the order of the elements $A$ in the class $C_{i}$, i.e. the smallest
integer ($>0$) for which the equation $A ^{^{\circ} h_{C_{i}}}= \mathbb{1}$
holds. Furthermore the table contains one representative for each
class $C_{i}$ given as product of the generators $S$ and $T$ of the group.
 \label{chartabTprime}}
\end{minipage}
\end{center}
\end{table}
The generators $S$ and $T$ fulfill the relations:
\begin{equation}
S^{2}=\mathbb{R}, \;\; T^{3}=\mathbb{1}, \;\; (S T)^{3}=\mathbb{1}, \;\; \mathbb{R}^{2}=\mathbb{1},
\end{equation}
where $\mathbb{R}=\mathbb{1}$ in case of the odd-dimensional representation and
$\mathbb{R}=-\mathbb{1}$ for $2$, $2^{\prime}$ and $2 ^{\prime \, \prime}$ such that $\mathbb{R}$ commutes with
all elements of the group. From Table 1 we see that, beyond the center of the group, generated by the elements
$E$ and $\mathbb{R}$, there are other abelian subgroups: $Z_3$, $Z_4$ and $Z_6$.
In particular, there is a $Z_4$ subgroup here denoted by $G_S$, generated by the element $TST^2$ and a $Z_3$ subgroup here
called $G_T$, generated by the element $T$. 
As we will see $G_S$ and $G_T$ are of great importance for the structure of our model.
Realizations of $S$ and $T$ for
$2$, $2^{\prime}$, $2^{\prime \, \prime}$ and $3$ can be found in the Appendix A and are taken from \cite{aranda}.
 
The multiplication rules of the representations are as follows:
 \be
 \begin{array}{l}
 1^a\otimes r^b=r^b\otimes 1^a=r^{a+b}~~~~~~~{\rm for}~r=1,2\\
 1^a\otimes 3=3\otimes 1^a=3\\
 2^a\otimes 2^b=3\oplus 1^{a+b}\\
 2^a\otimes 3=3\otimes 2^a=2\oplus 2'\oplus2''\\
3\otimes 3=3\oplus 3\oplus 1\oplus 1'\oplus 1''
 \end{array}
 \label{mult}
 \ee
where $a,b=0,\pm1$ and we have denoted $1^0\equiv1$, $1^{1}\equiv1'$, $1^{-1}\equiv1''$ and similarly for the doublet representations.
On the right-hand-side the sum $a+b$ is modulo 3.
The Clebsch-Gordan coefficients for the decomposition of product representations are shown in the Appendix A and were already calculated in \cite{aranda}.
Further synonyms of $T^{\prime}$ are Type 24/13 \cite{tprime} and $SL_{2} (F_{3})$ \cite{carr}.

%
\section{Outline of the model}
In this section we introduce our model and we illustrate its main features. 
We choose the model to be supersymmetric, which would help us when discussing the vacuum selection and the symmetry breaking pattern
of  $T'$. The model is required to be invariant under a flavour symmetry group $F=T'\times Z_3\times \Uu_{FN}$. The group factor $T'$ is the one
responsible for the TB lepton mixing. The group $T'$ is unable to produce all the necessary mass suppressions for the fermions.
These suppressions originate in part from a spontaneously broken U(1)$_{FN}$, according to the original FN proposal. Finally, the $Z_3$ factor helps 
in keeping separate
the contributions to neutrino masses and to charged fermion masses, and it is an important ingredient in the vacuum alignment analysis.
The fields of the model, together with their transformation properties under the flavour group, are listed in Table 2. 
 \begin{table}[!ht]
                \begin{math}
                \begin{array}{|c||c|c|c|c||c|c|c|c|c|c||c|c|c|c|c|c|}
                    \hline
                    &&&&&&&&&&&&&& &&\\[-9pt]
                    \text{Field} & l & e^c & \mu^c & \tau^c &  D_q
                     & D_u^c & D_d^c
                    & q_3 & t^c & b^c  &  h_{u,d} & \phit & \phis & \xi, \xit & \eta & \xipp \\[10pt]
                    \hline
                    &&&&&&&&&&&&&&&&\\[-9pt]
                    T' & 3 & 1 & \onepp & \onep & \twopp & \twopp & \twopp & 1 & 1 & 1 & 1 & 3 & 3 & 1 &  \twop & \onepp \\[3pt]
                    \Zt & \om & \om^2 & \om^2 & \om^2 & \om & \om^2 & \om^2 & \om & \om^2 & \om^2 & 1 & 1 & \om & \om  & 1 & 1 \\[3pt]
                    \Uu_{FN} & 0 & 2n & n & 0 & 0 & n & n & 0 & 0 & n & 0 & 0 & 0 & 0 & 0 & 0 \\[3pt]
                    \hline
                \end{array}
                \end{math}
            \caption{ The transformation rules of the fields under the
            symmetries associated to the groups \rm{$T'$}, \rm{$\Zt$} and
            \rm{$\Uu_{FN}$}.  We denote $D_q=(q_1,q_2)^t$ where
        $q_1=(u,d)^t$ and $q_2=(c,s)^t$ are the
        electroweak $\SUd$ doublets of the first two generations, $D_u^c=(u^c,c^c)^t$ and
        $D_d^c=(d^c,s^c)^t$.  $D_q$, $D_u^c$ and $D_d^c$ are doublets of $T'$.
         $q_3=(t,b)^t$ is the electroweak $\SUd$ doublet of the third generation.
         $q_3$, $t^c$ and $b^c$ are all singlets
         under $T'$.}
            \end{table}

\vskip 0.5cm
\subsection{Pattern of symmetry breaking}
 The most important feature of our model is the pattern of symmetry breaking of the flavour group $T'$. We will see that, at the leading order, 
 $T'$ is broken down to the subgroup $G_S$, generated by the element $TST^2$, in the neutrino sector and to the subgroup $G_T$, generated by $T$, 
 in the charged fermion sector. This pattern of symmetry breaking is achieved dynamically and corresponds to a local minimum of the scalar potential
 of the model. This result is already sufficient to understand the predicted pattern of fermion mixing angles. 
 Indeed, given  the $T'$ assignment of the matter fields displayed in Table 2 and the explicit expressions
 of the generators $S$ and $T$ for the various representations (see Appendix A), specific mass textures are obtained
 from the requirement of invariance under $TST^2$ or $T$. For instance, neutrinos are in a triplet of $T'$ and the element $TST^2$ in the triplet representations
 is given by:
\be
TST^2=\dfrac{1}{3}
\left(\begin{array}{ccc}
-1 & 2 & 2 \\
2 & -1 & 2 \\
2 & 2 & -1 \\
\end{array}\right)~~~.
\ee
The most general mass matrix for neutrinos invariant under $G_S$, in arbitrary units, is given by:
\be
 m_\nu=
 \left(
 \begin{array}{ccc}
 a+c&-b/3-c+d&-b/3\\
-b/3-c+d&c&a-b/3\\
-b/3&a-b/3&d
 \end{array}
 \right)
 \label{t1}
 \ee
where $a$, $b$, $c$ and $d$ are arbitrary parameters.
Similarly, the most general mass matrices for charged fermions invariant under $G_T$ have the following structure:       
\be
 m_e=
 \left(
 \begin{array}{ccc}
 \times & 0 & 0\\
 0 & \times & 0\\
 0 & 0 & \times
 \end{array}
 \right)
 \label{t2}
 \ee       
\be
 m_{u,d}=
 \left(
 \begin{array}{ccc}
 0 & 0 & 0\\
 0 & \times & \times\\
 0 & \times & \times
 \end{array}
 \right)
 \label{t3}
 \ee       
where a cross denotes a non-vanishing entry. The lepton mixing originates completely from $m_\nu$ and,
with an additional requirement, reproduces the TB scheme. This additional requirement is the condition $c=d$,
which is not generically implied by the invariance under $G_S$. In our model the  
fields that break $T'$ along the $G_S$ direction are a triplet $\varphi_S$ and an invariant singlet $\xi$. 
There are no further scalar singlets, transforming as $1'$ or $1''$ that couple to the neutrino sector.
We will see in a moment that due to this restriction our model gives rise to a particular version of the neutrino mass matrix in eq. (\ref{t1}), 
where $c=d=2b/3$, which implies directly a TB mixing.
It is interesting to note that, while the requirement of $G_T$ invariance implies a diagonal mass matrix in the
charged lepton sector, this is not the case for the quark sector, due to the different $T'$ assignment. At the leading order, in both up and down sectors,
we get mass matrices with vanishing first row and column, eq. (\ref{t3}). Moreover, the element 33 of both mass matrices is larger than the other elements,
since it is invariant under the full $T'$ group, not only the $G_T$ subgroup. The other non-vanishing elements carry a suppression factor originating from the
breaking of $T'$ down to $G_T$. This pattern of quark mass matrices, while not yet fully realistic, is however encouraging, since it reproduces correctly masses
and mixing angle of the second and third generations. As we will see, the textures in eqs. (\ref{t1},\ref{t2},\ref{t3}) are modified by subleading effects.
These effects are sufficiently small to keep the good feature of the leading order approximation, and large enough to provide a realistic description of the quark sector.

Fermion masses are generated by the superpotential $w$: 
\be
w=w_l+w_q+w_d
\label{fullw}
\ee 
where $w_l$ is the term responsible for the Yukawa interactions 
in the lepton sector, $w_q$ is the analogous term for quarks and $w_d$ is the term responsible for the vacuum alignment,
which will be discussed in the next section. We will consider the expansion of $w$ in inverse powers
of the cut-off scale $\Lambda$ and we will write down only the first non-trivial terms of this expansion. This will provide 
a leading order approximation, here analyzed in detail. Corrections to this approximation are produced by higher dimensional operators
contributing to $w$, which will be studied subsequently. As we will see in section 4, at the leading order, 
the scalar components of the supermultiplets $\varphi_T$, $\varphi_S$, $\xi$, $\tilde{\xi}$, $\eta$ and $\xi''$ develop VEVs
\be
\langle\varphi_S\rangle=(v_S,v_S,v_S)~~~,~~~~~\langle\xi\rangle=u~~~,~~~~~\langle\tilde{\xi}\rangle=0~~~,
\label{love1}
\ee
\be
\langle\varphi_T\rangle=(v_T,0,0)~~~,~~~~~\langle\eta\rangle=(v_1,0)~~~,~~~~~\langle\xi''\rangle=0~~~.
\label{love2}
\ee
These VEVs can be very large, much larger than the electroweak scale. In section 4 we will see that it is reasonable to choose:
\be
\dd\frac{VEV}{\Lambda}\approx \lambda^2~~~,
\label{vevratio}
\ee
where VEV stands for the generic non-vanishing VEV in eq. (\ref{love1},\ref{love2}).
Since the ratio in eq. (\ref{vevratio}) represents the typical expansion parameter when including higher dimensional operators, the choice in eq. (\ref{vevratio})
keeps all the leading order results stable, up to correction of relative order $\lambda^2$.
There is a neat misalignment in flavour space between $\langle\varphi_T\rangle$, $\langle\eta\rangle$ and $\langle\varphi_S\rangle$:
$\langle\varphi_T\rangle=(v_T,0,0)$, $\langle\eta\rangle=(v_1,0)$ and $\langle\xi''\rangle=0$ break $T'$ down to the subgroup $G_T$, while $\langle\varphi_S\rangle=(v_S,v_S,v_S)$
breaks $T'$ down to the subgroup $G_S$. It is precisely this misalignment the origin of the mass textures in eqs. (\ref{t1},\ref{t2},\ref{t3}).

A certain freedom is present in our formalism and this can lead to
models that are physically equivalent though different at a superficial level, when comparing VEVs or mass matrices. One source of freedom
is related to the possibility of working with different bases for the generators $S$ and $T$.  Another source of freedom is related to the fact that
vacua that break $T'$ are degenerate and lie in orbits of the flavour group.
For instance, when we say that the set of VEVs in eq. (\ref{love2}) breaks $T'$ leaving invariant the $Z_3$ subgroup generated by $T$,
VEVs obtained from this set by acting with elements of $T'$ are degenerate and
they preserve other $Z_3$ subgroups of $T'$.
Both these sources of freedom can lead to mass matrices 
different than those explicitly shown in eqs. (\ref{t1},\ref{t2},\ref{t3}).
It is however easy to show that the different ``pictures'' are related by field redefinitions and the physical properties of the system, such as the mass eigenvalues and the physical mixing angles, are always the same.
Thus it is not restrictive to work in a particular basis and to choose a single representative VEV configuration, as we will do in the following.
\vskip 0.3cm

\subsection{Leptons}

Lepton masses are described by $w_l$, given by:
\be
\begin{array}{cl}
w_l=&y_e e^c (\varphi_T l)h_d/\Lambda+y_\mu \mu^c (\varphi_T l)'h_d/\Lambda+
y_\tau \tau^c (\varphi_T l)''h_d/\Lambda+ \\[5pt]
&(x_a\xi+\tilde{x}_a\tilde{\xi}) (ll)h_u h_u/\Lambda^2+x_b (\varphi_S ll)h_u h_u/\Lambda^2+{\rm h.o.}
\end{array}
\label{wlplus}
\ee
where here and in the following formulae ${\rm h.o.}$ stands for higher dimensional operators.
After electroweak symmetry breaking, $\langle h_{u,d}\rangle=v_{u,d}$, given the specific orientation of $\langle\varphi_T\rangle\propto (1,0,0)$,
$w_l$ gives rise to diagonal mass terms for charged leptons:
\be
m_l=\frac{y_l v_T}{\sqrt{2}\Lambda} v_d~~~~~~~(l=e,\mu,\tau)~~~.
\ee
The $T'$ symmetry, as was the case for $A_4$, is unable to produce the required hierarchy among $m_e$, $m_\mu$ and $m_\tau$ and,
to this purpose, we make use of an additional spontaneously broken U(1)$_{FN}$ flavour symmetry. We introduce a new supermultiplet $\theta$,
carrying U(1)$_{FN}$ charge $-1$ and neutral under all other symmetries. Its non-vanishing $VEV$, $\langle\theta\rangle/\Lambda<1$ breaks U(1)$_{FN}$
and provides an expansion parameter for charged lepton masses.
We also assign U(1)$_{FN}$ charges $(2 n, n)$ to the fields $(e^c, \mu^c)$. All other lepton fields are taken neutral under this abelian symmetry.
In this way $y_\tau\approx 1$, $y_\mu\approx(\langle\theta\rangle/\Lambda)^{n}$, $y_e\approx(\langle\theta\rangle/\Lambda)^{2n}$ and
the mass hierarchy can be reproduced by choosing
\be
\left(\dd\frac{\langle\theta\rangle}{\Lambda}\right)^{n}\approx\lambda^2~~~.
\ee
All the information about lepton mixing angles is encoded in the neutrino mass matrix that can be easily evaluated from eq. (\ref{wlplus}) :
\be
m_\nu=\frac{v_u^2}{\Lambda}\left(
\begin{array}{ccc}
a+2 b/3& -b/3& -b/3\\
-b/3& 2b/3& a-b/3\\
-b/3& a-b/3& 2 b/3
\end{array}
\right)~~~,
\label{mnu0}
\ee
where
\be
a\equiv x_a\frac{u}{\Lambda}~~~,~~~~~~~b\equiv x_b\frac{v_S}{\Lambda}~~~.
\label{ad}
\ee
Notice that $m_\nu$ is not the most general mass matrix invariant under $G_S$, eq. (\ref{t1}). This is due to the absence of fields transforming
as $1'$ and $1''$ under $T'$, developing non-vanishing VEVs and directly contributing to $m_\nu$.
The neutrino mass matrix is diagonalized by the transformation:
\be
U^T m_\nu U =\frac{v_u^2}{\Lambda}{\tt diag}(a+b,a,-a+b)~~~,
\ee
with
\be
U=\left(
\begin{array}{ccc}
\sqrt{2/3}& 1/\sqrt{3}& 0\\
-1/\sqrt{6}& 1/\sqrt{3}& -1/\sqrt{2}\\
-1/\sqrt{6}& 1/\sqrt{3}& +1/\sqrt{2}
\end{array}
\right)~~~.
\ee
Therefore the TB mixing of eq. (\ref{tbm}) is reproduced, at the leading order.
For the neutrino masses we obtain:
\bea
|m_1|^2&=&\left[-r+\frac{1}{8\cos^2\Delta(1-2r)}\right]
\Delta m^2_{atm}\nn\\
|m_2|^2&=&\frac{1}{8\cos^2\Delta(1-2r)}
\Delta m^2_{atm}\nn\\
|m_3|^2&=&\left[1-r+\frac{1}{8\cos^2\Delta(1-2r)}\right]\Delta m^2_{atm}~~~,
\label{lospe}
\eea
where $r\equiv \Delta m^2_{sol}/\Delta m^2_{atm}
\equiv (|m_2|^2-|m_1|^2)/(|m_3|^2-|m_1|^2)$,
$\Delta m^2_{atm}\equiv|m_3|^2-|m_1|^2$ 
and $\Delta$ is the phase difference between
the complex numbers $a$ and $b$. For  $\cos\Delta=-1$, we have a neutrino spectrum close to hierarchical:
\be
|m_3|\approx 0.053~~{\rm eV}~~~,~~~~~~~
|m_1|\approx |m_2|\approx 0.017~~{\rm eV}~~~.
\ee
\begin{figure}[h!]
\centerline{\psfig{file=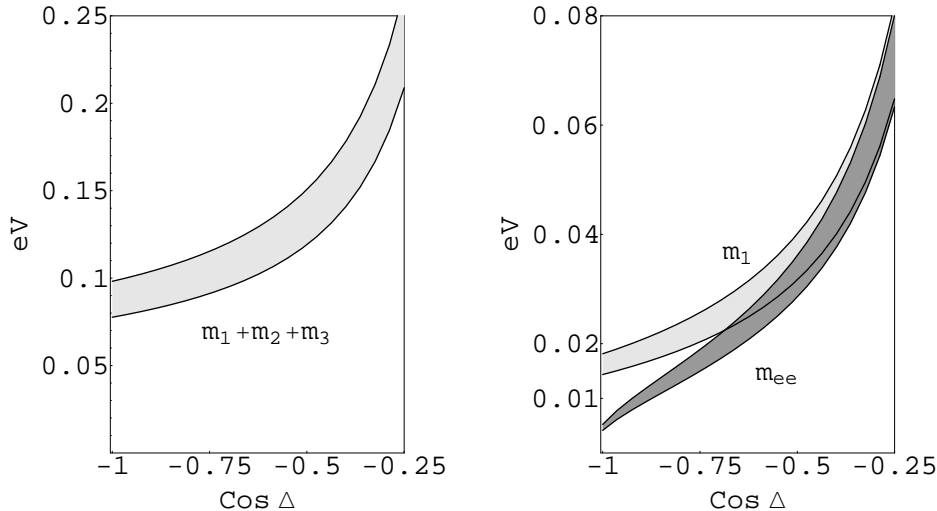,width=0.8\textwidth}}
\caption{On the left panel, sum of neutrino masses versus $\cos\Delta$, the phase difference between $a$ and $b$. On the right panel, the lightest neutrino mass, $m_1$ and the mass combination $m_{ee}$ versus $\cos\Delta$.
To evaluate the masses, the parameters $\vert a\vert$ and $\vert b\vert$
have been expressed in terms of $r\equiv \Delta m^2_{sol}/\Delta m^2_{atm}
\equiv (|m_2|^2-|m_1|^2)/(|m_3|^2-|m_1|^2)$ and
$\Delta m^2_{atm}\equiv|m_3|^2-|m_1|^2$. The bands have been obtained by 
varying $\Delta m^2_{atm}$ in its 3$\sigma$ experimental range, 0.0020 eV $\div$ 0.0032 eV. There is a negligible sensitivity to the variations of $r$
within its current 3 $\sigma$ experimental range, and we have
realized the plots by choosing $r=0.03$.}
\end{figure} 
In this case the sum of neutrino masses is about $0.087$ eV.
If $\cos\Delta$ is accidentally small, the neutrino spectrum becomes
degenerate. The value of $|m_{ee}|$, the parameter characterizing the 
violation of total lepton number in neutrinoless double beta decay,
is given by:
\be
|m_{ee}|^2=\left[-\frac{1+4 r}{9}+\frac{1}{8\cos^2\Delta(1-2r)}\right]
\Delta m^2_{atm}~~~.
\ee
For $\cos\Delta=-1$ we get $|m_{ee}|\approx 0.005$ eV, at the upper edge of
the range allowed for normal hierarchy, but unfortunately too small
to be detected in a near future.
Independently from the value of the unknown phase $\Delta$
we get the relation:
\be
|m_3|^2=|m_{ee}|^2+\frac{10}{9}\Delta m^2_{atm}\left(1-\frac{r}{2}\right)~~~,
\ee
which is a prediction of our model. In Figure 1 we have plotted the neutrino masses predicted by the model.
All the results listed above coincide with those obtained in the $A_4$ models of ref. \cite{af1,af2,afl}, at the leading order in the VEV expansion.
We will see that, when higher order effects are included, these results are modified by terms of relative order VEV/$\Lambda$. 
If this parameter is of order $\lambda^2$, as assumed in eq. (\ref{vevratio}), we have a stable TB mixing in the lepton sector. 

\subsection{Quarks}
The contribution to the superpotential in the quark sector is given by        
 \beq
            \vspace{0.15cm}
            \begin{array}{cl}
            w_q=\phantom{+}&y_t\left(t^c q_3\right)h_u+y_b\left(b^c
                    q_3\right)h_d\;+\\[3pt]
                &y_1(\phit D_u^c D_q) h_u/\La+
            y_5(\phit D_d^c D_q) h_d/ \La\;+\\[3pt]
                &y_2~\xipp(D_u^c D_q)' h_u/ \La+
            y_6~\xipp(D_d^c D_q)' h_d / \La\;+\\[3pt]
                & \{ y_3~  t^c (\eta D_q) +y_4~ (D_u^c \eta )q_3 \} h_u/\La+
            \{ y_7~ b^c (\eta D_q) +y_8~ (D_d^c \eta )q_3 \} h_d/\La\;+\\[3pt]
            &+{\rm h.o.}
            \label{wmq}
            \end{array}
            \vspace{0.1cm}
            \eeq
Observe that the supermultiplets $\varphi_S$, $\xi$ and $\tilde{\xi}$, which control the neutrino mass matrices, do not couple to the quark sector,
at the leading order. Conversely, the supermultiplets $\varphi_T$, $\eta$ and $\xi''$, which give masses to the charged fermions, do not couple
to neutrinos at the leading order. This separation is partly due to the discrete $Z_3$ symmetry, described in Table 2. 
By recalling the VEVs of eq. (\ref{love1},\ref{love2}), we can write down the mass matrices for the up and down quarks:
            \begin{gather}
            m_u=\left(\begin{array}{ccc}
                                0 & 0 & 0 \\
                                0 & y_1 \vt/\La & y_4  v_1/\La \\
                                0 & y_3 v_1 / \La & y_t \\
                        \end{array}\right)v_u+...\\[6pt]
            m_d=\left(\begin{array}{ccc}
                                0 & 0 & 0 \\
                                0 & y_5 \vt / \La & y_8 v_1/\La \\
                                0 & y_7 v_1/\La & y_b \\
                        \end{array}\right)v_d+...
\label{qmassm}
            \end{gather}
where dots stand for higher order corrections. These mass matrices are the most general ones that are invariant under $G_T$, see eq. (\ref{t3}).
The following quark masses and mixing angles are predicted, at the leading order:
\be
\begin{array}{c}
m_u=0~~~,~~~~~~~m_c\approx y_1 v_u v_T/\Lambda~~~,~~~~~~~m_t\approx y_t v_u\\[5 pt]
m_d=0~~~,~~~~~~~m_s\approx y_5 v_d v_T/\Lambda~~~,~~~~~~~m_b\approx y_b v_d\\[5 pt]
V_{us}=0~~~,~~~~~~~V_{ub}=0~~~,~~~~~~~V_{cb}\approx \left(\dd\frac{y_7}{y_b}-\dd\frac{y_3}{y_t}\right) \dd\frac{v_1}{\Lambda}
\end{array}
\ee
If the dimensionless coefficients $y_b$ and $y_5$ are of the same order of magnitude, the ratio $m_s/m_b$ is correctly reproduced since it is approximately given by $v_T/\Lambda$ and this ratio has been chosen of order $\lambda^2$, see eq. (\ref{vevratio}). To reproduce $m_c/m_t$ a further suppression is needed. This can be achieved by the usage of the U(1)$_{FN}$ symmetry by assigning a charge $n$ to the $T'$ doublet $D_u^c$. In this way the parameters $y_1$, $y_2$ and $y_4$ are of order $(\langle\theta\rangle/\Lambda)^{n}\approx\lambda^2$, and
\be
m_c/m_t\approx \dd\frac{y_1 v_T}{y_t \Lambda}\approx \left(\dd\frac{\langle\theta\rangle}{\Lambda}\right)^{n} \dd\frac{v_T}{\Lambda}\approx\lambda^4~~~.
\ee
The mass of the top quark is expected to be of the order of the VEV $v_u$ and then $y_t$ is naturally of $\cO(1)$. On the other hand, the mass of the bottom quark should be smaller than $m_t$ and be close to $m_\tau$. This is easily achieved by exploiting the U(1)$_{FN}$ symmetry, assigning  a charge $n$ to the $T'$ singlet $b^c$: as a result $y_b$ is of order $(\mean{\theta}/\Lambda)^n\approx\lambda^2$. 
Note that as a consequence in our model small and moderate values of $\tan\beta=v_u/v_d$ are preferred.
To preserve the ratio $m_s/m_b$, we need $y_5\approx \lambda^2$ as well that can originate if also the $T'$ doublet $D_d^c$ has a U(1)$_{FN}$ charge $n$. With this assignment of the FN charges, all the Yukawa couplings in the down quark sector are suppressed and of order of $\lambda^2$. 
Concerning the mixing, the element $V_{cb}$ is of order $v_1/\Lambda\approx\lambda^2$, reproducing the experimental data well.
Masses and mixing angles are however still unrealistic, since $m_u/m_c$, $m_d/m_s$, $V_{ub}$ and $V_{us}$ are vanishing, at this level.
We will see that all these parameters can be generated by higher order corrections,
in particular those affecting the VEVs in eq. (\ref{love1},\ref{love2}).
%
%
\section{The vacuum alignment at the leading order}
In this section we will discuss the minimization of the scalar potential of the model. 
To this purpose we should complete the definition of the superpotential $w$ by specifying
the last term in eq. (\ref{fullw}). This is the term responsible for the spontaneous symmetry breaking
of $T'$ and it includes a new set of fields, the `driving' fields, whose transformation properties are 
shown in Table 3. 
\begin{table}[!h]
\centering
\begin{tabular}{|c||c|c|c|c|c|}
\hline
&&&&&\\[-9pt]
Field & $\varphi ^{0} _{T}$ & $\varphi ^{0} _{S}$ & $\xi ^{0}$ & $\eta ^{0}$& $ \xi ^{\prime \, 0}$ \\
&&&&&\\[-9pt]
\hline
&&&&&\\[-9pt]
$T^{\prime}$ & 3 & 3 & 1 & $2 ^{\prime \, \prime}$ & $1 ^{\prime}$\\[3pt]
$Z_{3}$ & 1 & $\omega$ & $\omega$ & 1&1\\ [3pt]
\hline
\end{tabular}
\caption{The transformation rules of the driving fields under the
            symmetries associated to the groups $T'$, $Z_3$.}
            \label{flavoncharges}
\end{table}
Along the lines described in ref. \cite{af2}, we also exploit a U(1) R-symmetry
of the theory. All `matter' supermultiplets, those describing quarks and leptons,
have R-charge 1, while all supermultiplets that will develop a VEV, like $h_{u,d}$, $\varphi_{S,T}$,
$\xi$, $\tilde{\xi}$, $\xi''$, $\eta$ and $\theta$, have vanishing R-charges.
Driving fields have R-charge 2 and their contribution to the superpotential reads:
\begin{eqnarray}\nonumber
w_d &=& M \, (\varphi ^{0} _{T} \, \varphi_{T}) + g \, (\varphi ^{0} _{T} \, \varphi_{T} \, \varphi_{T}) + g_{7} \, \xi ^{\prime \, \prime} \, (\varphi ^{0} _{T} \, \varphi_{T})^{\prime} + g_{8} \, (\varphi ^{0} _{T} \, \eta \, \eta)\\ \nonumber
&+& g_{1} \, (\varphi ^{0} _{S} \, \varphi_{S} \, \varphi_{S}) + g_{2} \, \tilde{\xi} \, (\varphi ^{0} _{S} \, \varphi _{S})\\ \nonumber
&+& g_{3} \, \xi ^{0} \, (\varphi_{S} \, \varphi_{S}) + g_{4} \, \xi^{0} \, \xi^{2} + g_{5} \, \xi^{0} \, \xi \, \tilde{\xi} + g_{6} \, \xi^{0} \, \tilde{\xi} ^{2}\\ \nonumber
&+& M_{\eta} \, (\eta \, \eta^{0}) + g_{9} \, (\varphi_{T} \, \eta \, \eta^{0})\\
&+& M_{\xi} \, \xi ^{\prime \, \prime} \, \xi ^{\prime \, 0} + g_{10} \, \xi ^{\prime \, 0} \, (\varphi _{T} \, \varphi _{T}) ^{\prime \, \prime}+{\rm h.o.}
\end{eqnarray}
We start by analyzing the scalar potential in the supersymmetric limit. We look for a supersymmetric vacuum as the solution to the equations:
\begin{eqnarray}\nonumber
\frac{\partial w}{\partial \varphi^0_{S 1}}&=&g_2\tilde{\xi} {\varphi_S}_1+
\frac{2g_1}{3}({\varphi_S}_1^2-{\varphi_S}_2{\varphi_S}_3)=0\nn\\
\frac{\partial w}{\partial \varphi^0_{S 2}}&=&g_2\tilde{\xi} {\varphi_S}_3+
\frac{2g_1}{3}({\varphi_S}_2^2-{\varphi_S}_1{\varphi_S}_3)=0\nn\\
\frac{\partial w}{\partial \varphi^0_{S 3}}&=&g_2\tilde{\xi} {\varphi_S}_2+
\frac{2g_1}{3}({\varphi_S}_3^2-{\varphi_S}_1{\varphi_S}_2)=0\nn\\
\frac{\partial w}{\partial \xi^0}&=&
g_4 \xi^2+g_5 \xi \tilde{\xi}+g_6\tilde{\xi}^2
+g_3({\varphi_S}_1^2+2{\varphi_S}_2{\varphi_S}_3)=0
\label{neutrinomin}
\end{eqnarray}
\begin{eqnarray}\nonumber
\frac{\partial \, w}{\partial \, \varphi ^{0} _{T \, 1}} &=& M \, \varphi_{T \, 1} + \frac{2 \, g}{3} \, 
(\varphi ^{2} _{T \, 1} - \varphi _{T \, 2} \, \varphi _{T \, 3}) + g_{7} \, \xi ^{\prime \, \prime} \, \varphi_{T \, 2} + i \, g_{8} \, \eta_{1} ^{2}=0\\ \nonumber
\frac{\partial \, w}{\partial \, \varphi ^{0} _{T \, 2}} &=& M \, \varphi _{T \, 3} + \frac{2 \, g}{3} \, 
(\varphi ^{2} _{T \, 2} - \varphi _{T \, 1} \, \varphi _{T \, 3}) + g_{7} \, \xi ^{\prime \, \prime} \, 
\varphi_{T \, 1} + (1-i) \, g_{8} \, \eta_{1} \, \eta_{2} =0\\ \nonumber
\frac{\partial \, w}{\partial \, \varphi ^{0} _{T \, 3}} &=& M \, \varphi _{T \, 2} + \frac{2 \, g}{3} \, 
(\varphi ^{2} _{T \, 3} - \varphi _{T \, 1} \, \varphi_{T \, 2}) + g_{7} \, \xi ^{\prime \, \prime} \, \varphi_{T \, 3} + g_{8} \, \eta_{2} ^{2}=0\\
\label{chrgdmn}
\frac{\partial \, w}{\partial \, \eta ^{0} _{1}} &=& - M_{\eta} \, \eta_{2} + g_{9} \, ((1-i) \, \eta_{1} \, 
\varphi_{T \, 3} - \eta_{2} \, \varphi_{T \, 1})=0\\ \nonumber
\frac{\partial \, w}{\partial \, \eta ^{0} _{2}} &=& M_{\eta} \, \eta_{1} - g_{9} \, ((1+i) \, \eta_{2} \, 
\varphi_{T \, 2} + \eta_{1} \, \varphi_{T \, 1})=0\nonumber\\
\frac{\partial \, w}{\partial \, \xi ^{\prime \, 0}} &=& M _{\xi} \, \xi ^{\prime \, \prime} 
+ g_{10} \, (\varphi _{T \, 2} ^{2} + 2 \, \varphi _{T \, 1} \, \varphi _{T \,3})=0\nonumber
\end{eqnarray}
Concerning the first set of equations, eq. (\ref{neutrinomin}), there are flat directions in the SUSY limit. We can enforce $\langle\tilde{\xi}\rangle=0$ by adding to the scalar potential a soft SUSY breaking mass term
for the scalar field $\tilde{\xi}$, with $m^2_{\tilde{\xi}}>0$. In this case, in a finite portion of the parameter space, we find the solution
\begin{eqnarray}
\langle\tilde{\xi}\rangle&=&0\nn\\
\langle\xi\rangle&=&u\nn\\
\langle\varphi_S\rangle&=&(v_S,v_S,v_S)~~~,~~~~~~~~~v_S^2=-\frac{g_4}{3 g_3} u^2~~~
\label{solS}
\end{eqnarray}
with $u$ undetermined. By choosing $m^2_{\varphi_S},
m^2_\xi<0$, then $u$ slides to a large scale, which we assume to be eventually
stabilized by one-loop radiative corrections in the SUSY broken phase.
The VEVs in eq. (\ref{solS}) break $T'$ down to the subgroup $G_S$. 

Concerning the last six equations, eq. (\ref{chrgdmn}), by excluding the trivial solution where all VEVs vanish, in the SUSY limit we find three classes of solutions.
One class preserves the subgroup $G_S$, as for the set of VEVs given in eq. (\ref{solS}). It is characterized by $\langle\xi''\rangle\ne 0$ and
$\langle\eta\rangle=0$.
A representative VEV configuration in this class is:
\be
\langle \xi ^{\prime \, \prime} \rangle= - \frac{M}{g_{7}}~~~,~~~~~~~\langle \eta \rangle=(0,0)~~~,~~~~~~~\langle \varphi _T \rangle =(v_T,v_T,v_T)~~~,~~~~~~~~~v_T^2=\frac{M \, M_{\xi}}{3 \, g_{7} \, g_{10}}~~~.
\label{cl1}
\ee
The second class preserves a subgroup $Z_6$ generated by the elements $T$ and $\mathbb{R}$. It is characterized by $\langle\xi''\rangle=0$ and 
$\langle\eta\rangle=0$:
\be
\langle \xi ^{\prime \, \prime} \rangle =0~~~,~~~~~ \langle \eta\rangle =(0,0)~~~,~~~~~\langle \varphi_T\rangle =(v_T ,0,0)~~~,~~~~~~~
v_T=- \dd\frac{3M}{2g}
\label{cl21}
\ee
The third class preserves the subgroup $G_T$. It is characterized by $\langle\xi''\rangle=0$ and 
$\langle\eta\rangle\ne 0$:
\be
\begin{array}{c}
\langle \xi ^{\prime \, \prime} \rangle =0~~~,~~~~~\langle \eta\rangle = \pm(v_1 ,0)~~~,~~~~~\langle \varphi_T\rangle =(v_T,0,0)\\[10 pt]
v_1=\dd\frac{1}{g_9 \, \sqrt{3 \, g_8}} \, \sqrt{i \, (2 \, M_{\eta} ^{2} \, g + 3 \, M \, M_{\eta} \, g_9)}~~~,~~~~~~~v_T=\dd\frac{M_{\eta}}{g_{9}}~~~.
\label{cl22}
\end{array}
\ee
The three sets of minima in eqs. (\ref{cl1}), (\ref{cl21}) and (\ref{cl22}) are all degenerate in the SUSY limit and we will simply choose the one in eq. (\ref{cl22}). We have checked that, 
by adding soft masses $m_{\xi''}^2>0$, $m_\eta^2<0$,  the desired vacuum is selected as the absolute minimum, thus reproducing the result in eq. (\ref{love1},\ref{love2}). In summary, we have shown that
the VEVs in eqs. (\ref{love1},\ref{love2}) represent a local minimum of the scalar potential of the theory in a finite portion of the parameter space, without any ad hoc relation among the parameters of the theory.
As we will see in the next section, these VEVs will be slightly perturbed by higher order corrections induced by higher dimensional operators contributing to the `driving' potential $w_d$.
Such corrections will be important to achieve a realistic mass spectrum in the quark sector. Finally, concerning the numerical values of the VEVs, 
radiative corrections typically stabilize $u$ and $v_S$ well below the cut-off scale $\Lambda$. Similarly, mass parameters in the superpotential $w_d$
can be chosen in such a way that $v_1$ and $v_T$ are below $\Lambda$. It is not unreasonable to assume that all the VEVs are of the same order of magnitude:
\be
VEV\approx \lambda^2 \Lambda~~~.
\ee
%
%
\section{Higher order corrections}
The inclusion of higher order corrections is essential in our model. First of all, from these corrections we hope to achieve a realistic mass spectrum in the
quark sector. The leading order result is encouraging, but quarks of the first generation are still massless at this level and there is no mixing allowing 
communication between the first generation and the other ones. Moreover we should check that the higher order corrections do not spoil the leading order results. 
At the leading order there is a neat separation between the scalar fields giving masses to the neutrino sector and those giving masses to the charged fermion sector.
As a result the $T'$ flavour symmetry is broken down in two different directions in the two sectors:
neutrino mass terms are invariant under the subgroup $G_S$, while the charged fermion mass terms are invariant under 
the subgroup $G_T$. It is precisely this misalignment the source of the TB lepton mixing. Such a sharp separation is not expected to survive
when higher dimensional operators are included and this will cause the breaking of the subgroup $G_S$ $(G_T)$ in the neutrino (charged fermion) sector. 
It is important to check that this further breaking does not modify too much the misalignment achieved at the leading order and that the TB mixing remains stable.

The corrections are induced by higher dimensional operators, compatible with all the symmetries of our model, that can be included in the superpotential $w$,
thus providing the next terms in a $1/\Lambda$  expansion. It is convenient to discuss separately the higher order contributions to $w_l$, $w_q$ and $w_d$.
\subsection{Corrections to $w_l$}
The leading operators giving rise to $m_l$ are of order $1/\Lambda$ (see eq. (\ref{wlplus})). At order $(1/\Lambda)^2$ there are no new structures
contributing to $m_l$. Indeed, at this order the new invariant operators are:
\be
(f^c l\varphi_T\varphi_T) h_d/\Lambda^2~~~,~~~~~~~
(f^c \ell\eta\eta) h_d/ \La^2~~~,~~~~~~~
(f^c \ell\xipp\phit) h_d/ \La^2~~~,~~~~~~~(f^c=e^c,\mu^c,\tau^c)
\ee
Either they vanish because $\langle\xipp\rangle=0$ or they
replicate the leading-order pattern. 
The leading operators contributing to $m_\nu$ are of order $1/\Lambda^2$ (see eqs. (\ref{wlplus},\ref{mnu0},\ref{ad})).
At the next order we have two operators that vanish due to $\langle\xipp\rangle=0$:
\be
\xipp\xi (\ell\ell)^{\prime}h^2_u/ \La^3~~~,~~~~~
\xipp (\phis \ell\ell )^{\prime}h^2_u/ \La^3~~~,
\ee
and three new operators, whose contribution to $m_\nu$, after symmetry breaking, 
cannot be absorbed by a redefinition of the parameters $x_{a,b}$:
\be
(\varphi_T\varphi_S)' (ll)'' h_uh_u/\Lambda^3~~~,~~~~~~~(\varphi_T\varphi_S)'' (ll)' h_uh_u/\Lambda^3~~~,~~~~~
\xi (\varphi_T ll) h_uh_u/\Lambda^3~~~.
\label{hbc}
\ee
In addition to the above operators, there are also those obtained by replacing $\xi$ with $\tilde{\xi}$: they do not contribute at this order
due to the vanishing VEV of $\tilde{\xi}.$
The combined effects of these operators and of the corrections to the vacuum alignment of eqs. (\ref{love1},\ref{love2}) were discussed in ref. \cite{af2}.
Lepton masses and mixing angles are modified by terms of relative order $\lambda^2$. This correction is within the 1$\sigma$ experimental error
for $\theta_{12}$ and largely within the current uncertainties of $\theta_{23}$ and $\theta_{13}$. From the experimental view point, a small non-vanishing
value $\theta_{13}\approx \lambda^2$ and a deviation from $\pi/4$ of order $\lambda^2$ of
$\theta_{23}$, are both close to the reach of the next generation of neutrino experiments and will provide a valuable test of this model.

\subsection{Corrections to $w_q$}
The higher dimensional operators contributing to $w_q$ fall into two classes: those depending on
some of the fields $\varphi_S$, $\xi$, $\tilde{\xi}$ and those not depending on any of these fields.
Leading operators in the first class are necessarily cubic in $\varphi_S$, $\xi$, $\tilde{\xi}$,
due to the $Z_3$ symmetry given in Table 2. Therefore they will induce corrections to the quark mass
matrices at maximum of order $1/\Lambda^3$. These corrections might be at maximum of order $\lambda^6$ for the third row of $m_u$, while  at maximum of order $\lambda^8$ for the other rows of $m_u$ and for $m_d$, due to the additional suppression due to the FN mechanism. As a result these corrections are completely negligible, with the exception of contributions of order $\lambda^8$ to the 11 entry of both the quark mass matrices, which will arise anyway at this order in $\lambda$ also through other effects.
Operators in the second class depend only on the $T'$ breaking fields
$\varphi_T$, $\eta$ and $\xi''$, whose VEV pattern, eq. (\ref{love1},\ref{love2}), leaves invariant the subgroup $G_T$. 
Since the quark mass matrices shown in eq. (\ref{qmassm}) are already the
most general mass matrices invariant under this subgroup, any higher order operator
contributing to $w_q$ and leaving $T$ invariant after spontaneous breaking will predict the same textures of eq. (\ref{qmassm})
and its effect can be absorbed in a redefinition of the leading order coefficients.
Therefore we do not need to consider explicitly the new, next-to-leading, operators
contributing to the quark masses: either their effects are negligible or they can be absorbed
in a redefinition of the existing parameters. 
In the quark sector all the effects modifying the leading order results come from the corrections
to the VEVs in eq. (\ref{love1},\ref{love2}).
\subsection{Correction to $w_d$ and to the vacuum alignment}
Finally we are left with the corrections to the VEVs due to higher dimensional operators contributing to $w_d$.
We detail the discussion of this issue in the Appendix B. Here we only give the results. 
All the leading order operators in $w_d$ are of dimension three. After inclusion of a complete set of operators of dimension four,
the leading order VEVs
\be
\begin{array}{c}
\langle\varphi_T\rangle=(v_T,0,0)~~~,~~~~~\langle\varphi_S\rangle=(v_S,v_S,v_S)~~~\\[5pt]
\langle\xi\rangle=u~~~,~~~~~\langle\tilde{\xi}\rangle=0~~~,~~~~~
\langle\eta\rangle=(v_1,0)~~~,~~~~~\langle\xi''\rangle=0~~~
\end{array}
\label{lovevsbis}
\ee
are shifted into
\be
\begin{array}{c}
\langle\varphi_T\rangle=(v_T+\delta v_{T1},\delta v_{T2},\delta v_{T3})~~~,~~~~~\langle\varphi_S\rangle=
(v_S+\delta v_{S1},v_S+\delta v_{S2},v_S+\delta v_{S3})~~~\\[5pt]
\langle\xi\rangle=u~~~,~~~~~\langle\tilde{\xi}\rangle=\delta\tilde{u}~~~,~~~~~
\langle\eta\rangle=(v_1+\delta v_1,\delta v_2)~~~,~~~~~\langle\xi''\rangle=\delta u''~~~,
\end{array}
\label{hovevs}
\ee
where all the corrections $\delta v_{Ti}$, $\delta v_{Si}$, $\delta\tilde{u}$, $\delta u''$ and $\delta v_i$ are of order
$1/\Lambda$ and, given the large number of input parameters, they can be considered as mostly independent.
Since we typically have $VEV/\Lambda\approx \lambda^2$ at the leading order, we expect 
\be
\dd\frac{\delta VEV}{\Lambda}\approx \lambda^4~~~,
\ee
in a finite portion of the parameter space.
%
\section{Quark masses and mixing angles}
As explained in section 5,
the main effect of higher order corrections to quark masses comes from the modified VEVs, eq. (\ref{hovevs}).
When we insert these VEVs in $w_q$, eq. (\ref{wmq}), we get new quark mass matrices:
\be
\begin{array}{rcl}
            m_u&=&\left(\begin{array}{ccc}
                                i y_1 \delta v_{T2}/\Lambda +...& (1-i) y_1 \delta v_{T3}/2\Lambda+y_2\delta u''/\Lambda& -y_4 \delta v_2/\La \\
                                (1-i) y_1 \delta v_{T3}/2\Lambda-y_2\delta u''/\Lambda & y_1 v_T/\La & y_4 v_1/\La \\
                                - y_3 \delta v_2/\La & y_3 v_1 / \La & y_t \\
                        \end{array}\right)v_u\\[6pt]
            m_d&=&\left(\begin{array}{ccc}
 i y_5 \delta v_{T2}/\Lambda+... & (1-i) y_5 \delta v_{T3}/2\Lambda+y_6\delta u''/\Lambda& -y_8 \delta v_2/\La \\
                                (1-i) y_5 \delta v_{T3}/2\Lambda-y_6\delta u''/\Lambda & y_5 v_T/\La & y_8 v_1/\La \\
                                - y_7 \delta v_2/\La & y_7 v_1 / \La & y_b \\
                       \end{array}\right)v_d
\label{hoqm}
\end{array}
\ee
where we have redefined $v_T+\delta v_{T1}\to v_T$ and $v_1+\delta v_1\to v_1$ and the dots in the 11 entry of $m_u$ and $m_d$ 
stand for additional contributions from higher dimensional operators. 
Not all the available parameter space is suitable to correctly reproduce the masses and the mixing angles of the first generation quarks.
Indeed, by recalling that, generically, we expect $\delta VEV/VEV \approx \lambda^2$, we see that in this regime we would obtain, up to small corrections
\be
\dd\frac{m_u}{m_c}= \dd\frac{m_d}{m_s}= \dd\frac{\delta v_{T2}}{v_T}\approx \lambda^2~~~,
\ee
which is not correct in the up sector. To overcome this difficulty we assume that the correction $\delta v_{T2}$ is somewhat smaller than its natural value:
\be
\dd\frac{\delta v_{T2}}{v_T}\approx \lambda^4~~~.
\label{ass1}
\ee
This brings the up quark mass in the correct range but depletes too much the down quark mass. To get the appropriate mass for the down quark we assume that the dimensionless coefficient $y_6$ is not of its natural order $\lambda^2$, but enhanced by a factor
$1/\lambda$:
\be
y_6\approx\dd\lambda^2\frac{1}{\lambda}\approx\dd\lambda~~~.
\label{ass2}
\ee
We cannot justify the two assumptions in eqs. (\ref{ass1},\ref{ass2}) within our approach, where, in the absence of a theory for the higher-order terms, we have allowed for the
most general higher-order corrections. From our effective lagrangian approach, they should be seen as two moderate tunings that we need in order to 
get up and down quark masses.
To summarize, in our parameter space we naturally have $y_t$ and  $y_3$ of order one, while all other dimensionless parameters, with the exception of $y_6$, are of order $\lambda^2$, due to the FN mechanism. Concerning the VEVs, we can naturally accommodate
$VEV/\Lambda\approx\delta VEV/VEV\approx \lambda^2$, with the exception of $\delta v_{T2}$.
Within the restricted region of the parameter space where the two relations in eqs. (\ref{ass1},\ref{ass2}) are approximately valid, 
the quark mass matrices have the following structures:
\be
\begin{array}{rcl}
            m_u&=&\left(\begin{array}{ccc}
                                \lambda^8 & \lambda^6 & \lambda^6 \\
                                \lambda^6 & \lambda^4 & \lambda^4 \\
                                \lambda^4 & \lambda^2 & 1 \\
                        \end{array}\right)v_u\\[3mm]
            m_d&=&\left(\begin{array}{ccc}
                                \lambda^6 & \lambda^3 & \lambda^4 \\
                                \lambda^3 & \lambda^2 & \lambda^2 \\
                                \lambda^4 & \lambda^2 & 1 \\
                        \end{array}\right)\lambda^2v_d~~~.
\label{qmtextures}
\end{array}
\ee
By diagonalizing the matrices in eq. (\ref{hoqm}) with standard perturbative techniques we obtain: 
\be
\begin{array}{ll}
m_u\approx \left\vert y_1 v_u \left\{i\dd\frac{\delta v_{T2}}{\Lambda}-\left[\left(\dd\frac{1-i}{2}\right)^2 \dd\frac{\delta v_{T3}^2}{v_T\Lambda}-\dd\frac{y_2^2}{y_1^2} \dd\frac{\delta u''^2}{v_T\Lambda}\right]\right\}+...\right\vert\;,&m_d\approx \left\vert v_d \dd\frac{y_6^2}{y_5}\dd\frac{\delta u''^2}{v_T\Lambda}\right\vert\;,\\[15 pt]
m_c\approx \left\vert y_1 v_u \dd\frac{v_T}{\Lambda}\right\vert+O(\lambda^6)\;,&m_s\approx \left\vert y_5 v_d \dd\frac{v_T}{\Lambda}\right\vert+O(\lambda^6)\;,\\[10 pt]
m_t\approx \left\vert y_t v_u\right\vert+O(\lambda^4)\;, &m_b\approx \left\vert y_b v_d\right\vert+O(\lambda^6)~~.
\end{array}
\ee
For the mixing angles, we get:
\be
\begin{array}{l}
V_{ud}\approx V_{cs}\approx 1+O(\lambda^2)~~~~~~~~~~V_{tb}\approx 1\\[10 pt]
V_{us}^*\approx -V_{cd}\approx-\dd\frac{y_6}{y_5}\dd\frac{\delta u''}{v_T}-
\left[\left(\dd\frac{1-i}{2}\right)\dd\frac{\delta v_{T3}}{v_T}-\dd\frac{y_2}{y_1}\dd\frac{\delta u''}{v_T}\right]+O(\lambda^3)\\[10 pt]
V_{ub}^*\approx-\left(\dd\frac{y_7}{y_b}-\dd\frac{y_3}{y_t}\right) \left\{\dd\frac{\delta v_2}{\Lambda}+\dd\frac{v_1}{v_T}
\left[\left(\dd\frac{1-i}{2}\right)\dd\frac{\delta v_{T3}}{\Lambda}-\dd\frac{y_2}{y_1}\dd\frac{\delta u''}{\Lambda}\right]     \right\}\\[10 pt]
V_{cb}^*\approx-V_{ts}\approx\left(\dd\frac{y_7}{y_b}-\dd\frac{y_3}{y_t}\right) \dd\frac{v_1}{\Lambda}+O(\lambda^4)\\[15 pt]
V_{td}\approx-\dd\frac{y_6}{y_5} \left(\dd\frac{y_7}{y_b}-\dd\frac{y_3}{y_t}\right) \dd\frac{v_1\delta u''}{v_T \Lambda}+
\left(\dd\frac{y_7}{y_b}-\dd\frac{y_3}{y_t}\right) \dd\frac{\delta v_2}{\Lambda}
\end{array}
\ee
where, when not explicitly indicated, the relations include all terms up to $O(\lambda^4)$.
In the previous expressions, where all the quantities are generically complex, is possible to remove all phases except the 
one carried by the combination $(y_7/y_b-y_3/y_t)\delta v_2/\Lambda$ which enters $V_{ub}$ and $V_{td}$ at the order $\lambda^4$.
Notice that in our model $V_{ub}$ is of order $\lambda^4$ whereas $V_{td}$ is of order $\lambda^3$. In the
Wolfenstein parametrization of the mixing matrix, this corresponds to a combination $\rho+i\eta$ of order $\lambda$, which is phenomenologically
viable. Notice that quark masses and mixing angles are all determined within their correct order of magnitudes and enough parameters are present to
fit the data. Moreover, despite the large number of parameters controlling the quark sector, 
our model contains a well-known \cite{gst} non-trivial relation between masses and mixing angles:
\be
\sqrt{\dd\frac{m_d}{m_s}}=\left\vert V_{us}\right\vert+O(\lambda^2)~~~.
\label{p1}
\ee
Due to the approximate unitarity relation $V_{td}+V_{us}^* V_{ts}+V_{ub}^*=0$ and due to the fact that $V_{ub}$ is of order $\lambda^4$ 
in our model, from the relation in eq. (\ref{p1}) we also get:
\be
\sqrt{\dd\frac{m_d}{m_s}}=\left\vert\dd\frac{V_{td}}{V_{ts}}\right\vert+O(\lambda^2)~~~.
\label{p2}
\ee
These relations coincide well with the data: from \cite{pdg} we have $\sqrt{m_d/m_s}=0.213\div 0.243$, $\vert V_{us}\vert=0.2257\pm0.0021$
and $\vert V_{td}/V_{ts}\vert=0.208^{+0.008}_{-0.006}$. Unfortunately, the theoretical errors affecting
eqs. (\ref{p1}) and (\ref{p2}), dominated respectively by the unknown $O(\lambda^2)$ term in $V_{us}$ and by the unknown $O(\lambda^4)$ term in
$V_{td}$, are of order $20\%$. For this reason, and for the large uncertainty on the ratio $m_d/m_s$, it is not possible to turn these predictions
into precise tests of the model.
It is interesting to compare our predictions to those of early models of quark masses based on U(2) or $T'$ flavour symmetries \cite{qrel}.
They also predict eq. (\ref{p2}), with a smaller theoretical error of order $\lambda^3$. 
Moreover, due to the characteristic two zero textures, in their early versions they predict $\sqrt{m_u/m_c}=\vert V_{ub}/V_{cb}\vert$,
which is off by approximately a factor two. In our model the mass of the up quark depends on additional free parameters,
that modify this wrong relation by a relative factor of order one.

%
%
\section{Conclusion}
We have built a SUSY model of fermion masses and mixing angles based on the flavour symmetry group $T'\times \Uu_{FN}\times Z_3$.
In our model the key role is played by the discrete group $T'$, the double covering of $A_4$.
In the lepton sector our model maintains all the good properties of the models based on the symmetry group $A_4$ \cite{ma1,ma2,af1,af2,afl}. 
Indeed, since the group $T'$ possesses the representations $1$, $1'$, $1''$ and $3$, in our model we can reproduce the construction
made in refs. \cite{af1,af2,afl} and we obtain the same results for lepton masses and mixing angles.
The main point of our whole paper is that the symmetry group $T'$ allows to achieve a realistic description of quark masses and mixing angles
without spoiling the results in the lepton sector.
To describe quarks, we make use of the doublet representations of $T'$, not available in $A_4$. As was done in previous models for quark masses 
based on $T'$ and on U(2) \cite{aranda,barbieri}, we accommodate the first two generations of quarks in doublets under $T'$, whereas the third generation
is kept invariant. Such an assignment has several advantages. First of all, it allows to generate a mass for the top quark at the renormalizable level,
whereas the masses of the bottom quark and of the tau lepton stem from operators suppressed by one
power of the cutoff scale $\Lambda$. In our model the latter two masses are naturally of the same order
of magnitude.
Moreover, this choice of quark representations does not necessarily imply diagonal quark mass matrices, 
thus overcoming one of the major difficulty in extending $A_4$ to the quark sector. 
At the leading order only quarks of the second and third generations acquire masses and we can consistently
describe $m_t$, $m_c$, $m_b$, $m_s$ and $V_{cb}$.
To this purpose we need a set of scalar fields coupled to quarks, transforming non-trivially under $T'$ and developing 
appropriate VEVs in order to break $T'$ along the desired direction, the one left invariant by the generator $T$. We have explicitly verified that our model possesses
these requirements. Masses and mixing angles of the first quark generation are produced via higher-order effects 
induced by higher dimensional operators, which are compatible with all the symmetry requirements of the model, but
are depleted by inverse powers of the cutoff scale $\Lambda$ with respect to the leading order contributions.
Therefore the minimization of the full scalar potential, including all non-leading effects is a central aspect of our model,
crucial for a correct description of light quarks. As a result of such a minimization, which we have detailed in the Appendix B,
we find that in the parameter space of our model, extended by the introduction of higher dimensional operators, there is enough room to fit all the quark data.
At the same time, some constraints remain and we get the two approximate relations $\sqrt{m_d/m_s}\approx \vert V_{us}\vert$ and 
$\sqrt{m_d/m_s}\approx \vert V_{td}/V_{ts}\vert$.
These are well confirmed experimentally within the predicted $O(\lambda^2)$ uncertainty. 

In the lepton sector the model predicts a nearly
TB lepton mixing \cite{hps}, in very good agreement with the data. As in previous models based on $A_4$, such a mixing pattern is produced 
by a special breaking of the $T'$ group. At the
leading order and in the charged lepton sector, $T'$ is broken down to  the subgroup generated by the element $T$. This implies a diagonal
mass matrix for $e$, $\mu$ and $\tau$, with an hierarchy induced by the U(1)$_{FN}$ component of the full flavour symmetry group.
All the mixing originates from the neutrino sector, where $T'$ breaks down to the subgroup generated by the element $TST^2$. The source of
the TB lepton mixing is precisely the misalignment in flavour space between neutrino and charged lepton mass matrices \cite{volkas}.
A very important point of our model is that all sources of flavour symmetry breaking, including those
pertaining to the quark sector, do not spoil the successful leading order result for the neutrino mixing angles. Indeed,
higher order corrections modify such a mixing pattern only by terms of relative order $\lambda^2$, $\lambda\approx 0.22$ being the
Cabibbo angle. Future oscillation neutrino experiments will test the model to this level of accuracy.

All together we have five relations: three for the lepton mixing angles,
with relative accuracy $\lambda^2$ and two for the quark mixing angles with relative accuracy $\lambda$.
In the absence of a theory concerning the origin of the higher order corrections, it seems difficult to improve these predictions from the theory side
and the most stringent test of the model is still represented by an accurate measurement of $\theta_{23}$ and $\theta_{13}$, at the $\lambda^2$ level.
Additional tests of the model can be searched for in the context of rare processes, both in leptonic and in hadronic transitions.
Depending on the type of SUSY breaking, an imprint of the assumed flavour structure might survive in the SUSY breaking sector
and it might give rise to specific signatures on which we hope to come back in a future work.

\section*{Acknowledgments}
This work was made possible by the very kind hospitality of Manfred Lindner at the TUM and 
at the Max-Planck-Institut f\"ur Kernphysik in Heidelberg. We all thank him also for useful discussions and for participating 
in the early stage of the project. We also acknowledge useful discussions with Isabella Masina.
We recognize that this work has been partly supported by the European Commission under contracts MRTN-CT-2004-503369 and MRTN-CT-2006-035505.
\vfill
\newpage
\section{Appendix A}

        The matrices $S$ and $T$ representing the generators depend on the
        representations of the group:
                    \begin{equation}
                        \begin{array}{ccccc}
                            1 &\qquad& S=1 &\quad& T=1\\
                            1^\prime &\qquad& S=1 &\quad& T=\omega\\
                            1^{\prime\prime} &\qquad& S=1 &\quad& T=\omega^2\\[10pt]
                            2 &\qquad& S=A_1 &\quad& T=\omega A_2\\
                            2^\prime &\qquad& S=A_1 &\quad& T=\omega^2 A_2\\
                            2^{\prime\prime} &\qquad& S=A_1 &\quad& T=A_2\\[10pt]
                            3 &\qquad& S=\dfrac{1}{3}\left(\begin{array}{ccc}
                                                            -1 & 2\omega & 2\omega^2 \\
                                                            2\omega^2 & -1 & 2\omega \\
                                                            2\omega & 2\omega^2 & -1 \\
                                                            \end{array}\right)
                                    &\quad& T=\left(\begin{array}{ccc}
                                                            1 & 0 & 0 \\
                                                            0 & \omega & 0 \\
                                                            0 & 0 & \omega^2 \\
                                                        \end{array}\right)
                        \end{array}\nonumber
                    \end{equation}
        where we have used the matrices

            \begin{equation}
            A_1=-\dfrac{1}{\sqrt{3}}\left(\begin{array}{cc}
                                                i & \sqrt2e^{i\pi/12} \\
                                                -\sqrt2e^{-i\pi/12} & -i \\
                                            \end{array}\right)\,\qquad
            A_2=\left(\begin{array}{cc}
                          \omega & 0 \\
                          0 & 1 \\
                        \end{array}\right)\;.\nonumber
            \end{equation}\\

        \indent We now report the multiplication rules
        between the various representations. In the following we
        use $\alpha_\imath$ to indicate the elements of the first
        representation of the product and $\beta_\imath$ to indicate
        those of the second representation. Moreover $a,b=0,\pm1$ and
        we denote $1^0\equiv1$, $1^1\equiv1^\prime$,
        $1^{-1}\equiv1^{\prime\prime}$ and similarly for the
        doublet representations. On the right-hand side the sum
        $a+b$ is modulo 3.\\
        \indent We start with all the
        multiplication rules which include the 1-dimensional
        representations:
            \[
                \begin{array}{l}
                1\otimes Rep=Rep\otimes1=Rep\quad\text{with $Rep$ whatever
                representation}\\[8pt]
                1^a\otimes1^b=1^b\otimes1^a=1^{a+b}\equiv\alpha\beta\\[8pt]
                1^a\otimes2^b=2^b\otimes1^a=2^{a+b}\equiv\left(\begin{array}{c}
                                                                                \alpha\beta_1 \\
                                                                                \alpha\beta_2 \\
                                                                        \end{array}\right)\\[-10pt]\\[8pt]

                1^\prime\otimes3=3=\left(\begin{array}{c}
                                            \alpha\beta_3 \\
                                            \alpha\beta_1 \\
                                            \alpha\beta_2\\
                                    \end{array}\right)
                \qquad1^{\prime\prime}\otimes3=3=\left(\begin{array}{c}
                                            \alpha\beta_2 \\
                                            \alpha\beta_3 \\
                                            \alpha\beta_1\\
                                    \end{array}\right)
                \end{array}
            \]

        The multiplication rules with the 2-dimensional
        representations are the following:
            \[
                \begin{array}{lc}
                2\otimes2=2^\prime\otimes2^{\prime\prime}=2^{\prime\prime}\otimes2^\prime=3\oplus1&\quad
                \text{with}\left\{\begin{array}{ll}
                                    3=\left(\begin{array}{c}
                                        \dfrac{1-i}{2}(\alpha_1\beta_2+\alpha_2\beta_1) \\
                                        i\alpha_1\beta_1 \\
                                        \alpha_2\beta_2
                                    \end{array}\right)\\
                                    1=\alpha_1\beta_2-\alpha_2\beta_1\\
                                    \end{array}
                            \right.\\
                \\[-8pt]
                2\otimes2^\prime=2^{\prime\prime}\otimes2^{\prime\prime}=3\oplus1^\prime&\quad
                \text{with}\left\{\begin{array}{ll}
                                    3=\left(\begin{array}{c}
                                        \alpha_2\beta_2\\
                                        \dfrac{1-i}{2}(\alpha_1\beta_2+\alpha_2\beta_1) \\
                                        i\alpha_1\beta_1
                                    \end{array}\right)\\
                                    1^\prime=\alpha_1\beta_2-\alpha_2\beta_1
                                    \end{array}
                            \right.\\
                \\[-8pt]
                2\otimes2^{\prime\prime}=2^\prime\otimes2^\prime=3\oplus1^{\prime\prime}&\quad
                \text{with}\left\{\begin{array}{ll}
                                    3=\left(\begin{array}{c}
                                        i\alpha_1\beta_1 \\
                                        \alpha_2\beta_2\\
                                        \dfrac{1-i}{2}(\alpha_1\beta_2+\alpha_2\beta_1)
                                    \end{array}\right)\\
                                    1^{\prime\prime}=\alpha_1\beta_2-\alpha_2\beta_1
                                    \end{array}
                            \right.\\
                \\[-8pt]
                2\otimes3=2\oplus2^\prime\oplus2^{\prime\prime}&\quad
                \text{with}\left\{\begin{array}{ll}
                                    2\;=\;\left(\begin{array}{c}
                                                (1+i)\alpha_2\beta_2+\alpha_1\beta_1 \\
                                                (1-i)\alpha_1\beta_3-\alpha_2\beta_1
                                            \end{array}\right)\\
                                    2^\prime=\;\left(\begin{array}{c}
                                                (1+i)\alpha_2\beta_3+\alpha_1\beta_2 \\
                                                (1-i)\alpha_1\beta_1-\alpha_2\beta_2
                                            \end{array}\right)\\
                                    2^{\prime\prime}=\left(\begin{array}{c}
                                            (1+i)\alpha_2\beta_1+\alpha_1\beta_3 \\
                                            (1-i)\alpha_1\beta_2-\alpha_2\beta_3
                                             \end{array}\right)
                                    \end{array}
                            \right.\\
                \\[-8pt]
                2^\prime\otimes3=2\oplus2^\prime\oplus2^{\prime\prime}&\quad
                \text{with}\left\{\begin{array}{ll}
                                    2\;\,=\left(\begin{array}{c}
                                                (1+i)\alpha_2\beta_1+\alpha_1\beta_3 \\
                                                (1-i)\alpha_1\beta_2-\alpha_2\beta_3
                                            \end{array}\right)\\
                                    2^\prime\,=\left(\begin{array}{c}
                                                (1+i)\alpha_2\beta_2+\alpha_1\beta_1 \\
                                                (1-i)\alpha_1\beta_3-\alpha_2\beta_1
                                            \end{array}\right)\\
                                    2^{\prime\prime}=\left(\begin{array}{c}
                                                (1+i)\alpha_2\beta_3+\alpha_1\beta_2 \\
                                                (1-i)\alpha_1\beta_1-\alpha_2\beta_2
                                            \end{array}\right)
                                    \end{array}
                            \right.\\
                \\[-8pt]
                2^{\prime\prime}\otimes3=2\oplus2^\prime\oplus2^{\prime\prime}&\quad
                \text{with}\left\{\begin{array}{ll}
                                    2\;\,=\left(\begin{array}{c}
                                                    (1+i)\alpha_2\beta_3+\alpha_1\beta_2 \\
                                                    (1-i)\alpha_1\beta_1-\alpha_2\beta_2
                                                \end{array}\right)\\
                                    2^\prime\,=\left(\begin{array}{c}
                                                    (1+i)\alpha_2\beta_1+\alpha_1\beta_3 \\
                                                    (1-i)\alpha_1\beta_2-\alpha_2\beta_3
                                                \end{array}\right)\\
                                    2^{\prime\prime}=\left(\begin{array}{c}
                                                    (1+i)\alpha_2\beta_2+\alpha_1\beta_1 \\
                                                    (1-i)\alpha_1\beta_3-\alpha_2\beta_1
                                                \end{array}\right)
                                    \end{array}
                            \right.\\
                \end{array}
            \]

        The multiplication rule with the 3-dimensional
        representations is
            \[
                3\otimes3=3_S\oplus3_A\oplus1\oplus1^\prime\oplus1^{\prime\prime}
            \]
        where
            \begin{gather*}
                3_S=\dfrac{1}{3}\left(\begin{array}{c}
                                         2\alpha_1\beta_1-\alpha_2\beta_3-\alpha_3\beta_2\\
                                         2\alpha_3\beta_3-\alpha_1\beta_2-\alpha_2\beta_1\\
                                         2\alpha_2\beta_2-\alpha_1\beta_3-\alpha_3\beta_1\\
                                \end{array}\right)\qquad
                3_A=\dfrac{1}{2}\left(\begin{array}{c}
                                         \alpha_2\beta_3-\alpha_3\beta_2\\
                                         \alpha_1\beta_2-\alpha_2\beta_1\\
                                         \alpha_3\beta_1-\alpha_1\beta_3\\
                                \end{array}\right)\\[10pt]
                \begin{array}{l}
                1\;\,=\alpha_1\beta_1+\alpha_2\beta_3+\alpha_3\beta_2\\
                1^\prime\,=\alpha_3\beta_3+\alpha_1\beta_2+\alpha_2\beta_1\\
                1^{\prime\prime}=\alpha_2\beta_2+\alpha_1\beta_3+\alpha_3\beta_1\;.
                \end{array}
            \end{gather*}

\section{Appendix B}
\noindent In this appendix we discuss the subleading terms of the superpotential $w_{d}$ 
and how they correct the VEV alignment.
We work along the lines of the Appendix B of \cite{af2}.

\noindent The VEVs are shifted from the values
\[
\langle \varphi_{S} \rangle= (v_{S},v_{S},v_{S}) \; , \;\; \langle \varphi_{T} \rangle= (v_{T},0,0) \; , \;\; \langle \eta \rangle = (v_{1},0) \; , \;\; \langle \xi \rangle = u \; , \;\; \langle \tilde{\xi}
\rangle = 0 \; , \langle \xi ^{\prime \, \prime} \rangle = 0 \;  
\]
\noindent to the values
\begin{eqnarray}\nonumber
&&\langle \varphi _{S} \rangle= (v_{S} + \delta v_{S \, 1},v_{S} + \delta v_{S \, 2}, v_{S} + \delta v_{S \, 3}) \; ,  \;\; \langle \varphi_{T} \rangle= (v_{T}+ \delta v_{T \, 1},\delta v_{T \, 2}, 
\delta v_{T \, 3}) \; , \;\;
\\ \nonumber
&& \langle \eta \rangle = (v_{1} + \delta v_{1}, \delta v_{2}) \; , 
\langle \xi \rangle = u \; , \langle \tilde{\xi}
\rangle = \delta \tilde{u} \; , \langle \xi ^{\prime \, \prime} \rangle = \delta u^{\prime \, \prime} 
\end{eqnarray}
\noindent where the corrections $\delta v_{T \, i}$, $\delta v_{S \, i}$, $\delta v_{i}$, 
$\delta \tilde{u}$ and $\delta u^{\prime \, \prime}$ are independent of each other. 
Note that there also might be a
correction to the VEV $u$, but we do not have to indicate this explicitly by the addition of
a term $\delta u$, since $u$ is undetermined at tree-level anyway.  

\noindent We change the notation in eq. (31) a bit by defining
\[
g_{3} \equiv 3 \, \tilde{g}_{3} ^{2} \;, \;\; g_{4} \equiv - \tilde{g}_{4} ^{2} \;\;\; \mbox{and}
\;\;\; g_{8} \equiv i \, \tilde{g} _{8} ^{2}
\]
\noindent such that the VEVs read
\[
v_{S}= \frac{\tilde{g}_{4}}{3 \, \tilde{g}_{3}} \, u \;, \;\; v_{T}= \frac{M_{\eta}}{g_{9}} \;\;\; 
\mbox{and} \;\;\; v_{1}=\frac{1}{\sqrt{3} \, \tilde{g}_{8} \, g_{9}} \, \sqrt{2 \, g\, M_{\eta} ^{2}
+ 3 \, g_{9} \, M \, M_{\eta}}
\]
where we have chosen the ``$+$'' sign for the VEV $v_{1}$. Apart from the subleading terms which
are already presented in \cite{af2} we get 17 other invariants which involve at least
one of the new fields $\eta _{1,2}$, $\xi ^{\prime \, \prime}$, $\eta^{0} _{1,2}$ and $\xi ^{\prime \, 
0}$:
\begin{equation}\nonumber
\Delta w_{d \, 2} = \frac{1}{\Lambda} \, \left( \sum \limits _{i=14} ^{18} t_{i} \, I_{i} ^{T}
 + \sum \limits _{i=13} ^{15} s_{i} \, I_{i} ^{S} + x_{4} \, I_{4} ^{X} + \sum \limits _{i=1} ^{4}
n_{i} \,  I_{i} ^{N} + \sum \limits _{i=1} ^{4} y_{i} \, I_{i} ^{Y} \right)
\end{equation}
with 
\small
\begin{eqnarray}\nonumber
I_{14} ^{T} &=& (\varphi ^{0} _{T} \, \varphi _{T}) ^{\prime \, \prime} \, \xi ^{\prime \, \prime} \, \xi ^{\prime \, \prime} 
= (\varphi ^{0} _{T \, 2} \, \varphi _{T \, 2} + \varphi ^{0} _{T \, 1} \, \varphi _{T \, 3} + \varphi ^{0} _{T \, 3}
\, \varphi _{T \, 1}) \, \xi ^{\prime \, \prime} \, \xi ^{\prime \, \prime} 
\\ \nonumber
I_{15} ^{T} &=& (\varphi _{T} \, \eta) \, (\varphi ^{0} _{T} \, \eta) = ((1+i) \, \varphi _{T \, 1} \, 
\eta _{2} + \varphi _{T \, 3} \, \eta _{1}) \, ((1-i) \, \varphi ^{0} _{T \, 2} \, \eta _{1} - 
\varphi ^{0} _{T \, 3} \, \eta _{2})
\\ \nonumber
&-& ((1-i) \, \varphi_{T \, 2} \, \eta_{1} - \varphi_{T \, 3} \, \eta_{2}) \, ((1+i) \, \varphi ^{0} _{T \, 1}
\, \eta_{2} + \varphi ^{0} _{T \, 3} \, \eta_{1})
\\ \nonumber
I_{16} ^{T} &=& (\varphi _{T} \, \eta)^{\prime} \, (\varphi ^{0}_{T} \, \eta) ^{\prime \, \prime}
= ((1+i) \, \varphi _{T \, 2} \, \eta _{2} + \varphi _{T \, 1} \, \eta _{1})\, ((1-i) \, \varphi ^{0} _{T \, 1}
 \, \eta _{1} - \varphi ^{0} _{T \, 2}  \, \eta _{2})
\\ \nonumber
&-& ((1-i) \, \varphi _{T \, 3} \, \eta _{1} - 
\varphi _{T \, 1} \, \eta _{2}) \, ((1+i) \, \varphi ^{0} _{T \, 3} \, \eta _{2} + \varphi ^{0} _{T \, 2} \, 
\eta_{1})
\\ \nonumber
I_{17} ^{T} &=& (\eta \, \xi ^{\prime \, \prime}) \, (\varphi ^{0} _{T} \, \eta)
= \xi ^{\prime \, \prime} \, (\eta _{1} \, ((1-i) \, \varphi ^{0} _{T \, 2} \, \eta _{1} - \varphi ^{0} 
_{T \, 3} \, \eta _{2}) - \eta _{2} \, ((1+i) \, \varphi ^{0} _{T \, 1} \, \eta _{2} + 
\varphi ^{0} _{T \, 3}\, \eta _{1}))
\\ \nonumber
I_{18} ^{T} &=& (\varphi _{T} \, \varphi _{T})_{S} \, (\varphi ^{0} _{T} \, \xi ^{\prime \, \prime})
= \frac{2}{3} \, ((\varphi ^{2} _{T \, 1} - \varphi _{T \, 2} \, \varphi _{T \, 3}) \, \varphi ^{0} _{T \, 2}
+ (\varphi ^{2} _{T \, 3} - \varphi _{T \, 1} \, \varphi _{T \, 2}) \, \varphi ^{0} _{T \, 1} + (
\varphi ^{2} _{T \, 2} - \varphi _{T \, 1} \, \varphi _{T \, 3}) \, \varphi ^{0} _{T \, 3}) \, \xi ^{\prime \, 
\prime}
\\ \nonumber
I_{13} ^{S} &=& (\varphi ^{0} _{S} \, \varphi _{S}) ^{\prime} \, \xi \, \xi ^{\prime \, \prime} 
= (\varphi ^{0} _{S \, 3} \,  \varphi _{S \, 3} + \varphi ^{0} _{S \, 1} \, \varphi _{S \, 2} 
+ \varphi ^{0} _{S \, 2} \, \varphi _{S \, 1}) \, \xi\, \xi ^{\prime \, \prime}
\\ \nonumber
I_{14} ^{S} &=&  (\varphi ^{0} _{S} \, \varphi _{S}) ^{\prime} \, \tilde{\xi} \, \xi ^{\prime \, \prime} 
= (\varphi ^{0} _{S \, 3} \,  \varphi _{S \, 3} + \varphi ^{0} _{S \, 1} \, \varphi _{S \, 2} 
+ \varphi ^{0} _{S \, 2} \, \varphi _{S \, 1}) \, \tilde{\xi} \, \xi ^{\prime \, \prime}
\\ \nonumber
I_{15} ^{S} &=& (\varphi _{S} \, \xi ^{\prime \, \prime}) \, (\varphi ^{0} _{S} \, \varphi _{S})_{S}
= \frac{1}{3} \, \xi ^{\prime \, \prime} \, (\varphi _{S \, 2} \, (2 \, \varphi ^{0} _{S \, 1} \, 
\varphi _{S \, 1} - \varphi ^{0} _{S \, 2} \, \varphi _{S \, 3} - \varphi ^{0} _{S \, 3} \, \varphi _{S \, 2}) 
\\ \nonumber
&+& \varphi _{S \, 3} \, (2 \, \varphi ^{0} _{S \, 2} \, \varphi _{S \, 2} - \varphi ^{0} _{S \, 1} \, \varphi _{S \, 3}
- \varphi ^{0} _{S \, 3} \, \varphi _{S \, 1})
+\varphi _{S \, 1} \, (2 \, \varphi ^{0} _{S \, 3} \, \varphi _{S \, 3} - \varphi ^{0} _{S \, 1} \, \varphi _{S \, 2}
- \varphi ^{0} _{S \, 2} \, \varphi _{S \, 1}))
\\ \nonumber
I_{4} ^{X} &=& (\varphi _{S} \, \varphi _{S}) ^{\prime} \, \xi ^{0} \, \xi ^{\prime \, \prime}
= (\varphi ^{2} _{S \, 3} + 2 \, \varphi _{S \, 1} \, \varphi _{S \, 2}) \, \xi ^{0} \, \xi ^{\prime \, \prime}
\end{eqnarray}
\noindent and furthermore the structures involving the driving fields $\eta ^{0} _{1,2}$ and
$\xi ^{0 \, \prime}$:
\begin{eqnarray}\nonumber
I_{1} ^{N} &=& (\varphi _{T} \, \varphi _{T}) \, (\eta ^{0} \, \eta)
= (\varphi ^{2} _{T \, 1} + 2 \, \varphi _{T \, 2} \, \varphi _{T \, 3}) \, (\eta ^{0} _{1} \, \eta_{2} 
- \eta ^{0} _{2} \, \eta_{1})
\\ \nonumber
I_{2} ^{N} &=& (\varphi _{T} \, \eta) \, (\eta ^{0} \, \varphi _{T})
= ((1+i) \, \varphi _{T \, 1} \, \eta _{2} + \varphi _{T \, 3} \, \eta _{1}) \, ((1-i) \, \eta ^{0} _{1}
\, \varphi _{T \, 1} - \eta ^{0} _{2} \, \varphi _{T \, 2}) 
\\ \nonumber
&-& ((1-i) \, \varphi _{T \, 2} \, \eta _{1} -
\varphi _{T \, 3} \, \eta _{2}) \, ((1+i) \, \eta ^{0} _{2} \, \varphi _{T \, 3} + \eta ^{0} _{1} \, 
\varphi _{T \, 2})
\\ \nonumber
I_{3} ^{N} &=& (\eta \, \xi ^{\prime \, \prime}) \, (\eta ^{0} \, \varphi _{T})
= \xi ^{\prime \, \prime} \, (\eta _{1} \, ((1-i) \, \eta ^{0} _{1} \, \varphi _{T \, 1} - \eta ^{0} _{2}
\, \varphi _{T \, 2}) - \eta _{2} \, ((1+i) \, \eta ^{0} _{2} \, \varphi _{T \, 3} + \eta ^{0} _{1}
\, \varphi _{T \, 2}))
\\ \nonumber
I_{4} ^{N} &=& (\eta \, \eta)_{3} \, (\eta ^{0} \, \eta)_{3}
= \frac{1}{2} \, (1+i) \, \eta _{1}^{2} \, (\eta ^{0}_{1} \, \eta _{2} + \eta ^{0} _{2} \, \eta _{1})
+ \eta _{2} ^{3} \, \eta ^{0}_{2} + (1+i) \, \eta _{1} ^{2} \, \eta _{2} \, \eta ^{0} _{1}
\\ \nonumber
I_{1} ^{Y} &=& (\varphi _{T} \, \varphi _{T}) \, \xi ^{\prime \, 0} \, \xi ^{\prime \, \prime}
= (\varphi ^{2} _{T \, 1} +  2 \, \varphi _{T \, 2} \, \varphi _{T \, 3}) \, \xi ^{\prime \, 0} \, 
\xi ^{\prime \, \prime}
\\ \nonumber
I_{2} ^{Y} &=& (\varphi _{T} \, \eta) ^{\prime} \, (\xi ^{\prime \, 0} \, \eta) ^{\prime \, \prime}
= ((1+i) \, \varphi _{T \, 2} \, \eta _{2} + \varphi _{T \, 1} \, \eta _{1}) \, \xi ^{\prime \, 0} \, \eta_{2}
- ((1-i) \, \varphi _{T \, 3} \, \eta _{1} - \varphi _{T \, 1} \, \eta _{2}) \, \xi ^{\prime \, 0} \, 
\eta _{1}
\\ \nonumber
I_{3} ^{Y} &=& (\varphi _{S} \, \varphi _{S}) ^{\prime \, \prime} \, \xi ^{\prime \, 0} \, \xi
= (\varphi ^{2} _{S \, 2} + 2 \, \varphi _{S \, 1} \, \varphi _{S \, 3}) \, \xi ^{\prime \, 0} \, \xi 
\\ \nonumber
I_{4} ^{Y} &=& (\varphi _{S} \, \varphi _{S}) ^{\prime \, \prime} \, \xi ^{\prime \, 0} \, \tilde{\xi}
= (\varphi ^{2} _{S \, 2} + 2 \, \varphi _{S \, 1} \, \varphi _{S \, 3}) \, \xi ^{\prime \, 0} \, \tilde{\xi} 
\end{eqnarray}

\noindent If we perform the analogous calculation as done in \cite{af2}, i.e. we only 
take into account terms which are at most linear in $\delta VEV$ and 
no terms of the order $O(\delta VEV/
\Lambda)$ where $\Lambda$ is the cutoff scale,
and plug in the VEVs $v_{T}$ and $v_{1}$, the equations take the form:

\small
\begin{eqnarray}\nonumber
&&\frac{\tilde{g}_{4} \, u^3}{3 \, \tilde{g}_{3} \, \Lambda} \, \left(t_{11}+\frac{\tilde{g}^2_{4}}{3
   \, \tilde{g}_{3}^2}\, \left( t_{6} +t_{7} + t_{8} \right)\, \right) + \frac{t_{3}}{\Lambda} \, v_{T}^3+(1-i) \, \frac{t_{16}}{\Lambda} \, v_{1}^2 \, v_{T}
- 2 \, v_{T} \, \left( \frac{2 \, g \, v_{T}}{3}+ M \right) \, \frac{\delta v_{1}}{v_{1}} \\ \label{deltavT1}
&&+ \left( M + \frac{4 \, g \, v_{T}}{3} \right) \, \delta v_{T \, 1} =0
\\ \label{deltavT2}
&& \frac{\tilde{g}_{4} \, u^3}{3 \, \tilde{g}_{3} \, \Lambda} \, \left(t_{11}+\frac{\tilde{g}^2_{4}}{3
   \, \tilde{g}_{3}^2}\, \left( t_{6} +t_{7} + t_{8} \right)\, \right) + \left( M -\frac{2 \, g \, v_{T}}{3} \right) \, 
\delta v_{T \, 2}=0
\\ \nonumber
&&\frac{\tilde{g}_{4} \, u^3}{3 \, \tilde{g}_{3} \, \Lambda} \, \left(t_{11}+\frac{\tilde{g}^2_{4}}{3
   \, \tilde{g}_{3}^2}\, \left( t_{6} +t_{7} + t_{8} \right)\, \right) + g_{7} \, v_{T} \, \delta u^{\prime \, \prime} + (1+i) \, v_{T} \,\left(
\frac{2 \, g \, v_{T}}{3}+ M \right) \,
   \frac{\delta v_{2}}{v_{1}}\\ \label{deltavT3}
&& + \left( M - \frac{2 \, g \, v_{T}}{3} \right) \, \delta v_{T \, 3}=0
\\  \label{deltavS1}
&& \left( \frac{9 \, \tilde{g} _{3} \, s _{10}}{\tilde{g}_{4}}
+ \frac{3 \, \tilde{g}_{4} \, s_{3}}{\tilde{g}_{3}}
+2 \, s_{6} \right) \, \frac{v_{T} \, u}{\Lambda} + 3 \, g_{2} \, \delta
   \tilde{u}+2 \, g_{1} \, \left( 2 \, \delta v_{S\, 1}- \delta v_{S \,2}-\delta v_{S \, 3} 
 \right)=0
\\ \label{deltavS2} 
&& \left( \frac{3 \, \tilde{g}_{4} \, s_{4}}{\tilde{g}_{3}}-s_{6} 
  - \frac{3}{2} \, s_{8} \right) \, \frac{v_{T} \, u}{\Lambda} + 3 \, g_{2} 
   \delta \tilde{u}  
   + 2 \, g_{1} \, \left( 2 \, \delta v_{S \, 2} - \delta v_{S \, 1} -\delta v_{S \, 3} \right)  =0
\\ \label{deltavS3}
&& \left( \frac{3 \, \tilde{g}_{4} \, s_{5}}{\tilde{g}_{3}}-s_{6}
   +\frac{3}{2} \, s_{8} \right) \, \frac{v_{T} \, u}{\Lambda} + 3 \, g_{2} \,
   \delta \tilde{u}+2 \, g_{1} \, \left( 2 \, \delta v_{S \, 3}- \delta
   v_{S\, 1}-\delta v_{S \, 2} \right)=0
\\ \label{deltau}
&& \frac{x_{2} \, v_{T} \, u}{3 \, \tilde{g}_{3} \, \Lambda}+ \frac{g_{5}}{\tilde{g}_{4}} \, \delta \tilde{u} +2 \,
   \tilde{g}_{3} \, \left( \delta v_{S \, 1} + \delta v_{S \, 2}+ \delta v_{S \, 3} \right) =0
\\ \label{deltav2}
&& v_{T} \, \delta v_{2}- \frac{1}{2}(1-i) \, v_{1} \, \delta v_{T \, 3}=0
\\ \label{deltav1}
&& -\frac{1}{2 \, \Lambda} \, (1+i) \, n_{4} \, v_{1}^2 + \frac{n_{1}}{\Lambda} \, v_{T}^2 + g_{9} \, \delta v_{T \, 1}=0
\\ \label{deltaupr2}
&& \frac{\tilde{g}_{4}^2 \, y_{3} \, u^3}{3 \, \tilde{g}_{3}^2 \, \Lambda}+M_{\xi} \, \delta u^{\prime \, \prime}+ 2 \,
   g_{10} \, v_{T} \, \delta v_{T \, 3}=0
\end{eqnarray}
\normalsize

\noindent As one can see the eqs. (\ref{deltavS1},\ref{deltavS2},\ref{deltavS3},\ref{deltau}) do 
not get any contributions from the terms of $\Delta w_{d \, 2}$, i.e. the shifts $\delta v_{S \, i}$
and $\delta u$ are the same as in the $A_4$ model. The eqs. (\ref{deltavT1},\ref{deltavT2},
\ref{deltavT3}) are also correlated to the analogous ones in the $A_4$ model. In order to see this one
has to set the couplings appearing in $\Delta w_{d \, 2}$ to zero and take into account that
$v_{T}= -\frac{3 \, M}{2 \, g}$ in the $A_4$ model such that 
$-2 \, v_{T} \, (\frac{2}{3} \, g \, v_{T} + M) \, \frac{\delta v_{1}}{v_{1}}$ vanishes and 
expressions like $(M + \frac{4 \, g \, v_{T}}{3}) \, \delta v_{T \, 1}$ are just 
$-M \, \delta v_{T \, 1}$. With this at hand the eqs. (\ref{deltavT1},\ref{deltavT2},
\ref{deltavT3}) fully coincide with the ones found in \cite{af2}.
The last three equations are not present in case of $A_4$ and they simply vanish, if the couplings 
and the VEVs of the fields only present in case of $T^{\prime}$ and not $A_4$ are set to zero.

Concerning the tuning we need in the parameter $\delta v_{T2}$, eq. (\ref{ass1}), it is easy to see that it is compatible with the structure of the above equations
and, at the same time, it has no consequences on the other shifts of the VEVs.

\end{document}